%% file: NewHiggsv3.tex
\documentclass[12pt]{article}
\usepackage{graphicx}
\usepackage[margin=1.25in]{geometry}
\usepackage[usenames,dvipsnames]{color}
\usepackage{url}
\usepackage{caption,subcaption, cite}
\usepackage[colorlinks = true,
            linkcolor = blue,
            urlcolor  = blue,
            citecolor = blue,
            anchorcolor = blue]{hyperref}

\newcommand\pubnumber{DESY 17-120 \\ 
 KEK Preprint 2017--22 \\
 SLAC--PUB--17129}
\newcommand\pubdate{August, 2017}

\def\SLAC{SLAC,
    Stanford University, Menlo Park, CA 94025, USA}

\def\kek{High Energy Accelerator Research Organization (KEK), Tsukuba,
  Ibaraki, JAPAN  }
\def\Tokyo{ICEPP, University of Tokyo, Hongo, Bunkyo-ku, Tokyo,
  113-0033, JAPAN}
\def\DESY{DESY,
    Notkestrasse 85,  22607 Hamburg, GERMANY}
\def\SNU{Dept. of Physics and Astronomy, Seoul National
  Univ.,
Seoul 08826, KOREA}

\def\Title#1{\begin{center} {\Large #1 } \end{center}}
\def\Author#1{\begin{center}{ \sc #1} \end{center}}

\def\submit#1{\begin{center}Submitted to {\sl #1} \end{center}}
\newcommand\pubblock{\rightline{\begin{tabular}{l} \pubnumber\\
         \pubdate \end{tabular}}}
\newenvironment{Abstract}{\begin{quotation} \begin{center}
                       ABSTRACT
     \end{center}\bigskip  }{\end{quotation}}

\def\submit#1{\begin{center}Submitted to {\sl #1} \end{center}}
\def\Acknowledgements{\bigskip  \bigskip \begin{center} \begin{large}
             \bf ACKNOWLEDGEMENTS \end{large}\end{center}}

\input mymacros.tex

\begin{document}
\begin{titlepage}
\pubblock

\vfill
\Title{Improved Formalism for Precision Higgs Coupling Fits}

\bigskip

\bigskip 

\Author{Tim Barklow$^a$, Keisuke Fujii$^b$, Sunghoon Jung$^{ac}$, Robert Karl$^d$,
 Jenny List$^d$, Tomohisa Ogawa$^b$, 
Michael E. Peskin$^{a}$, and 
  Junping Tian$^e$}

\bigskip

\begin{center} { \it
$^a$  \SLAC \\ 
$^b$  \kek \\
$^c$ \SNU \\ 
$^d$ \DESY \\ 
$^e$   \Tokyo }
\end{center}

\vfill

\begin{Abstract}
Future $\ee$ colliders give the promise of model-independent
determinations of the couplings of the Higgs boson.  In this paper,
we present an improved formalism for extracting Higgs boson couplings from 
$\ee$ data, based on the Effective Field Theory description of
corrections to the Standard Model.   We apply this formalism to give 
projections of Higgs coupling accuracies for stages of the
International Linear Collider and for other proposed $\ee$ colliders.
\end{Abstract}
\vfill
\submit{Physical Review D}
\vfill

\newpage
\tableofcontents
\end{titlepage}

\def\thefootnote{\fnsymbol{footnote}}
\setcounter{footnote}{0}
%



\section{Introduction}

One of the most important opportunities provided by future $\ee$
colliders is that of determining the couplings of the Higgs boson with
high precision and in a model-independent way.  By now, many studies
have made projections of the accuray with which Higgs couplings can be
determined at proposed $\ee$ 
colliders~\cite{ILCTDR,ILCWhite,ILCcase,TLEP,CEPC,CLIC,Lafaye}.  Most of these
studies are based on the $\kappa$ formalism, in which each  Standard
Model Higgs coupling is  multiplied by an independent factor
$\kappa_I$ and these factors are fit to expected measurements.  The
assumption is that introducing a large number of $\kappa_I$
parameters leads to coupling determinations with a great deal of
model-independence.

 An alternative method, described in \cite{SILH,SILHtwo},  takes a different point of
view.  This method begins from the assumption that the corrections to
the Standard Model (SM) due to new physics can be parametrized by the
addition of higher-dimension operators to the renormalizable
(dimension 4) SM  Lagrangian.  We have come to realize that this
approach is more correct in the way it  takes into   account the variety 
of effects that might arise
from new physics. In addition, it allows us to incorporate powerful
constraints from $SU(2)\times U(1)$ gauge invariance, and to make use 
of new observables that have not previously been considered in 
Higgs coupling fits.   In this paper, we will formalize this approach,
describe its advantages, and present projections of Higgs coupling
accuracy for future $\ee$ colliders based on this formalism.   Some recent
analyses with similar ingredients but different emphases can be
found in \cite{Gu,Ge,Sanz,Khanpour,DGGW,Vita,Chiu}. 

In particular, we make the assumption that the deviations from the
SM predictions for the Higgs couplings can be represented
by the addition of dimension-6 operators.   There are a large number
of possible dimension-6 operator coefficients---84 in all---but only a
manageable number of these play a role in the analysis of
 Higgs couplings.   We will
refer to the effective field theory (EFT) operator coefficients as
$c_J$, to distinguish them from the $\kappa_I$.  

 The EFT approach is
largely now adopted in the analysis of Higgs coupling data from the
LHC; see \cite{LHCEFT}.  However, the information on the EFT
coefficients that will come from future $\ee$ colliders will be much
more complete and specific.  In fact, it is  shown in some detail in
the accompanying paper \cite{BFJPT} that data from future $\ee$
colliders can determine, independently and without ambiguity, all of the
dimension-6 EFT coefficients that contribute directly to Higgs boson
production and decay processes at those colliders.  We can then use
the EFT approach to provide estimates of Higgs boson couplings that
are completely model-independent as long as the general framework of
the EFT is valid. 

  Because one EFT
operator can be exchanged for another by the use of the equations of
motion, there are several different conventions used for the $c_I$. In
this paper, we will use a variant of the ``Warsaw basis'' introduced
in \cite{Warsaw}.  The notation, and detailed formulae for the linear
deviations of the Higgs couplings, can be found in \cite{BFJPT}.   A
similar set of formulae in the ``SILH basis''~\cite{SILH,SILHtwo} can be found in
\cite{DGGW,CGP}.

 \section{Why is $\kappa_Z$ model-dependent?}

Let's get right to the point:  Why is a formalism based on $\kappa_I$
not model-independent?

   For Higgs decays to fermions, the corrections to the Standard Model
   are described phenomenologically by a single operator whose effect
   can be described by a $\kappa_I$ rescaling.  The same is true for
   decays to $gg$ or $\gamma\gamma$.  However, 
 for the Higgs coupling to
$WW$ and $ZZ$, this is not correct.   The EFT actually leads to two 
distinct structures.   We can represent the Higgs-$Z$ interaction
as parametrized by two coefficients  $\eta_Z$, $\zeta_Z$,
\beq
  \delta\L = (1+  \eta_Z) {m_Z^2\over v} h Z_\mu Z^{\mu} + \zeta_Z
  {h\over 2v} Z_{\mu\nu} Z^{\mu\nu} \ , 
\eeq{WLagrange}
where $Z_{\mu\nu}$ is the $Z$ field strength.  A similar formula can
be written for the Higgs-$W$ interaction. 
 The coefficients $\eta_Z$, $\eta_W$ multiply
vertices with the same form as the SM vertices, but the $\zeta_W$ and $\zeta_Z$
terms bring in a new interactions of a different form.   The $\eta$
and $\zeta$ parameters are derived from the EFT operator
coefficients $c_I$ in a way that we will discuss in a moment.

The $\zeta$ terms  involve the field
strengths of the vector fields, and so are  momentum-dependent.   The
effect of these terms depends on the momenta of the two vector bosons
and the extent to which these are off-shell.   For a 125 GeV Higgs
boson and the cross section at 250 GeV in the center of mass, we find,
to linear order in the corrections,
\beqa
    \sigma(\ee\to Zh) &=&  (SM) \cdot ( 1 + 2\eta_Z + (5.7)
                \zeta_Z )\CR
\Gamma(h\to WW^*) &=& (SM) \cdot (1 + 2\eta_W - (0.78)\zeta_W ) \CR
\Gamma(h\to ZZ^*) &=& (SM) \cdot (1 + 2\eta_Z - (0.50)\zeta_Z )  \ . 
\eeqa{etazetarel}
The coefficients
in front of the $\zeta$ terms come from integrals over the relevant
phase space for each process.  

 In weakly-coupled
extensions of the Higgs sector, including supersymmetry, the $\zeta$ 
coefficients
typically arise from loop diagrams and 
 have values of $10^{-3}$ or smaller, but in Little Higgs
and Randall-Sundrum models (without T-parity), these coefficients can
be present at 
the tree level and can be as large as
other new physics contributions~\cite{SILH,Elias}.  
 A simple $\kappa_I$ parametrization
cannot incorporate this degree of freedom.  Thus, we conclude, the
$\kappa_I$ formalism is not model-independent and cannot provide a
general basis for tests of  models of new physics effects on the Higgs
couplings against data.

The EFT analysis adds parameters to the standard $\kappa_I$ scheme,
but it also  has a compensatory advantage.  New physics corrections
can modify the relative size of the Higgs boson couplings to $Z$ and
$W$.
 In the  usual model-independent $\kappa_I$ analysis, this is
 accounted
by taking  $\kappa_Z$
and $\kappa_W$ to be independent parameters that can vary arbitrarily
with respect to one another.    In the EFT approach,
the largest contributions to the 
the parameters $\eta_Z$ and $\eta_W$ and to $\zeta_Z$ an $\zeta_W$ 
come from the same dimension-6
operator coefficients.  More explicitly, we find 
\beq
      \eta_W  = -\half c_H \qquad \eta_Z =    - \half c_H  - c_T \  ,
\eeq{etaWZ}
where $c_T$ is related the $T$ parameter of precision electroweak
analysis~\cite{PandT} and is constrained by that analysis to be very small.
The parameter  $c_H$ is an overall renormalization of the Higgs field
as discussed in \cite{SILH,Han,CraigMcC}.
  Similarly, with  $(c_w, 
s_w) = (\cos\theta_w,\sin\theta_w)$,  we find
\beqa
\zeta_W &=&  (8c_{WW} )\CR
\zeta_Z &=&   c_w^2 (8c_{WW}) + 2 s_w^2 (8c_{WB}) +
(s_w^4/c_w^2)(8c_{BB})\  ,
\eeqa{zetaWZ}
where  $c_{WW}$,  $c_{WB}$,
$c_{BB}$  are coefficients of dimension-6 operators with
 the squares of $SU(2)\times
U(1)$ field strengths. The parameters $c_{WB}$ and 
$c_{BB}$  are  strongly constrained by measurements  outside 
 the program of  $\ee$ measurements of  Higgs reactions.  Thus, in the EFT
 approach, the 
relative sizes of the two $Z$ and $W$ couplings are regulated by 
$SU(2)\times U(1)$ gauge invariance in a way that their relation can
be determined from data.
The overall effect is that we exchange the two parameters $\kappa_Z$, $\kappa_W$
for two parameters $\eta_Z$, $\zeta_Z$, with no new freedom for
$\eta_W$ and $\zeta_W$.

The structure of the Higgs couplings to $W$ and $Z$ is important to
resolve the trickiest and most subtle problem of Higgs coupling
analysis.  Experiments measure branching ratios, but models of new
physics predict the absolute strengths of Higgs couplings and, through
these, the Higgs partial widths.  To effectively compare theory and
experiment, it is necessary to find the absolute normalization of the
partial widths.   The conversion factor is the Higgs boson total
width.   This width, about 4~MeV in the SM, is too small to  be measured
directly at any proposed accelerator.  Rather, it must be extracted
from the fit to coupling constants. 

 In the literature, this is
typically done within the $\kappa$ framework 
 by assuming that the total cross section for $\ee\to
Zh$ and the partial width $\Gamma(h\to ZZ^*)$ are both proportional to
the parameter $\kappa_Z$.   This total cross section can be measured
by observing the recoil $Z$ at a fixed lab energy, independently of
the Higgs decay scheme.   This determines $\kappa_Z$ to high accuracy.
The Higgs width can then 
then be extracted from the ratio of measurable quantities
\beq
     { \sigma(\ee\to Zh)\over BR(h\to ZZ^*)} = 
    { \sigma(\ee\to Zh)\over \Gamma(h\to ZZ^*)/\Gamma_h } \ \sim\ 
    \Gamma_h \ ,
\eeq{usualGamma}
from which $\kappa_Z$ cancels out.   Since $BR(h\to ZZ^*)$
is small, about 3\% in the SM, this determination of $\Gamma_h$
suffers from low statistics, but, at least, it seems to be model-independent.

The presence of the $hZZ$ coupling proportional to $\zeta_Z$, ruins
this strategy.   We see from \leqn{etazetarel} that the numerator and
denominator
of \leqn{usualGamma} have completely different dependence on $\zeta_Z$,
even with a different sign.   To overcome this
problem, we need a separate method to determine the size of the
$\zeta_Z$ terms.   We will discuss this in the next section.

\section{Elements of a fit for $\eta_Z$ and $\zeta_Z$}

 There are a number of possible  methods to determine the size of the $\zeta$
parameters. In this section, we will highlight one particularly
powerful method, which is to make use of the angular distribution and
polarization asymmetries of the the reaction 
$\ee\to Zh$.  These observables have not previously been applied to Higgs coupling
analysis.  

The contributions to the $\ee\to
Zh$ cross section from the 
 the $\eta_Z$ and $\zeta_Z$ terms can be distinguished by their
 effects on these angular distributions and asymmetries.
The $\eta_Z$ terms lead to enhanced amplitudes for longitudinal
$Z$ polarization and to production at smaller values of $|\cos\theta|$, 
while the $\zeta_Z$ terms lead to equal production of the three $Z$
polarization states at higher values of $|cos\theta|$. At 250~GeV,
this is a relatively small effect, proportional to $(E_Z^2/m_Z^2 - 1)
= 0.47$, but it becomes larger at higher energy.  Second, the 
contribution
from the $\zeta_Z$ term is quite sensitive to beam polarization.
  Beam
polarization is straightforward to achieve at linear colliders but is
not projected for circular colliders, while circular collider designs
have higher luminosity at 250~GeV.  Then there is a certain
complementarity between these approaches.

\begin{figure}
\begin{center}
\includegraphics[width=0.70\hsize]{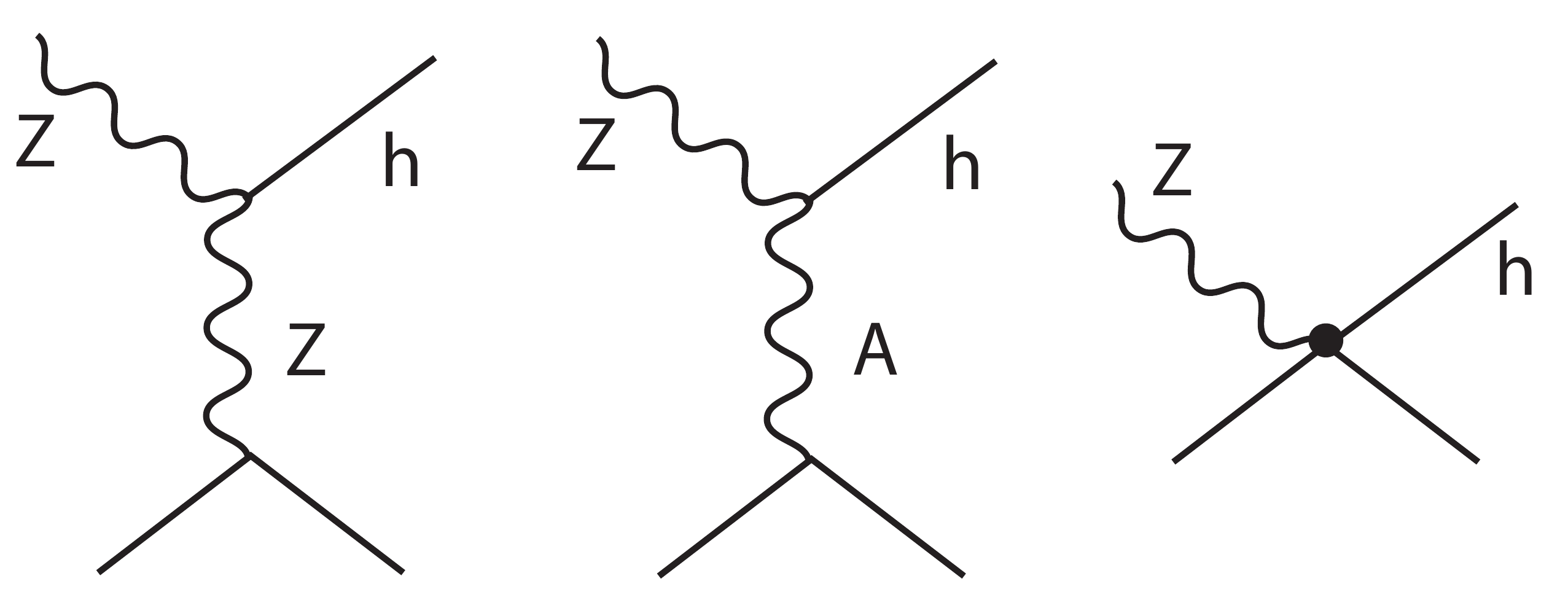}
\end{center}
\caption{Diagrams contributing to  the cross section for $\ee\to Zh$ in
  the EFT description of Higgs couplings.  The second and third
  diagrams are not present in the Standard Model.  They arise from the
  dimension-6 operators in the EFT.}
 \label{fig:ZAdiagrams}
\end{figure}

The polarization effect in the $\ee\to Zh$ cross section 
arises from the interference of $s$-channel  diagrams with $Z$ and
$A$; see Fig.~\ref{fig:ZAdiagrams}. 
  In addition to the $\zeta_Z$ term in the Higgs Lagrangian, the
dimension-6 operators in the EFT induce a term
\beq
    \delta \L =   \zeta_{AZ} {h\over v} A_{\mu\nu} Z^{\mu\nu} 
\eeqn
that mixes the $A$ and $Z$ field strengths.  The coefficient of this
term is related to the parameters already discussed by
\beq
    \zeta_{AZ} =   s_w c_w  (8c_{WW}) - {s_w\over c_w} (c_w^2 - s_w^2)
    (8c_{WB}) - {s_w^3\over c_w} (8c_{BB}) \ .
\eeqn
This produces the second diagram in Fig.~\ref{fig:ZAdiagrams}, which
is not present at tree level  in the SM.     There is also a third
contribution that has the form of a contact interaction.   This   third
diagram is parametrically enhanced by terms of order $s/m_Z^2$~\cite{Shaouly}.
However, the EFT coeffiecients that appear in this diagram $(c_{HL},
c_{HL}^\prime, c_{HE})$ are very strongly constrained by the analysis
of precision electroweak corrections and $\ee\to W^+W^-$, in a way
that compensates this enhancement~\cite{BFJPT}. We will
ignore this diagram in the simplified analysis presented in this section.
However, it does play a role in the more
complete analysis that we will describe in Section 4.

The sign of the  interference term between the $s$-channel $Z$ and $A$ diagrams
depends on the beam polarization.
 The $Z$ charge of the electron changes sign between
$e^-_L$ and $e^-_R$,  ($\half - s_w)^2 \to (-s_w^2)$), while the $A$
charge stays the same.   This leads to a near-cancellation
of the $\zeta_Z$ terms for $e^-_R$, while there is constructive
interference for $e^-_L$. 

To illustrate these effects, we carry out a simplified fit to the Higgs
couplings in the following framework:  The starting point is the table
of projected errors on the total $\ee\to Zh$ cross section given in
the 
Appendix.   These error estimates are based on full simulation studies
with the ILD detector model~\cite{ILDstudies,Ogawa}.  The estimates
are provided for each of two configurations of beam polarization, an
$L$ beam with electron and positron beam polarizations $(-0.8,+0.3)$
and an $R$ beam with electron and positron beam polarizations
$(+0.8,-0.3)$. These errors are essentially identical for the two beam
polarizations, and so we can apply them also for unpolarized beams. 
All of the analyses below are carried out at the
linearized level. 

A fit in the $\kappa$ framework would modify each Higgs couplings by 
\beq
             g_{hA\bar A} =   g_{hA\bar A} (1 + \delta\kappa_A) 
\eeqn
For reference, we carried out a $\kappa$ fit to this data from $\ee\to
Zh$ with 7
parameters:
\beq
  \delta \kappa_Z, \delta\kappa_W, \delta\kappa_b,
 \delta\kappa_c, \delta\kappa_g, 
 \delta\kappa_\tau, \delta\kappa_\mu \ . 
\eeqn

In addition, we allow branching ratios of the Higgs boson to
invisible and to non-invisible exotic decay modes. This modification
of the Standard Model is usually omitted in EFT fits, which
concentrate on the effects of heavy particles.  However, the search
for Higgs decays to light invisible and exotic particles is an
important part of the full $\ee$ program, and the possibility of such
decays adds an  uncertainty to the extraction of the Higgs boson total
width that should be accounted.
  To parametrize these possible exotic Higgs decays, we introduce
two additional parameters $\delta a_{inv}$ and $\delta a_{other}$.
The parameter $\delta a_{inv}$ is the fraction of all Higgs
  decays that go to completely invisible decay products, 
\beq
   \delta a_{inv} =   \Gamma(h\to \mbox{invis}) /\Gamma_{h,SM}  \ . 
\eeq{ainvis}
Similarly, $\delta a_{other}$ is the fraction of all Higgs decays that
do not fit into any standard category, or even into the category of 
invisible decays. 
It is extremely conservative to include the $\delta a_{other}$ parameter,
since, at an $\ee$ collider,  almost any exotic decay will be observed
and recognized as 
such~\cite{LiuExotics}.   But  this is the way that all previous
``model-independent'' Higgs coupling fits for $\ee$ colliders have been done.

The simplified fits in the EFT framework also uses 9
parameters.  These are the EFT parameters $c_H$ and $c_{WW}$, the EFT 
parameters that shift the Higgs couplings $b$, $c$, $g$, $\tau$, and
$\mu$,
and the $a_{inv}$ and $a_{other}$ parameters described above
\leqn{ainvis}~\cite{gapology}. In this simplified fit, $c_{WB}$ and $c_{BB}$ are set equal to
zero.  In the complete fit described below, these latter parameters
 are determined by
constraints from precision electroweak measurements, $\ee\to W^+W^-$, and
$\Gamma(h\to \gamma\gamma)$. 

\begin{table}
\begin{center}
\begin{tabular} {lcccccc}
 &  $\kappa$ fit &  angl. only & pol. only &  both &   full EFT fit \\
\hline
$g(hb\bar b)$ &            3.21  & 3.87 &   0.94  &  0.94  &   1.04          \\ 
$g(h c\bar c)$ &           3.52  &  4.19 &   1.73    &  1.73   &   1.79       \\ 
$g(h gg)$ &                   3.43  &  4.10  &   1.54  &  1.54 &   1.60     \\ 
$g(hWW)$ &                   3.31  & 3.77  &  0.46   & 0.45    &    0.65    \\ 
$g(h\tau\tau)$ &           3.25   &  3.91  &  1.07   & 1.07   &   1.16    \\ 
$g(hZZ)$ &                    0.36  &   3.51    & 0.45   &   0.44  &  0.66 \\ 
$g(h\mu\mu)$ &            13.1 & 14.7  &     12.8 &        12.8  &  5.53      \\ 
\hline
$g(hb\bar b)/g(hWW)$ &   0.85    &   0.92  & 0.83    &  0.83 &   0.82        \\ 
$g(hWW)/g(hZZ)$ &          3.29  &   0.26  & 0.02  & 0.02   &    0.07     \\ 
\hline
$\Gamma_h$ &                 6.53 &  7.64  &  2.20    & 2.18 &   2.38  \\ 
$\sigma(\ee\to Zh)$ &    0.72  & 0.80  &  0.72   & 0.70   &   0.70     \\ 
\hline
$BR(h\to inv)$ &        0.39 &    0.36    &   0.30    &  0.30 &  0.30    \\ 
$BR(h\to other)$ &  1.57   &      1.71  &   1.53 &   1.51  &   1.50 
  \\ 
\end{tabular}
\caption{Projected relative errors for Higgs boson couplings and other Higgs
observables, in \%, for the fits described in Section 3.   The first
column gives results of 
 a fit with simple $\kappa$ rescaling of the Standard Model Higgs
couplings. 
The next three  columns give results of fits to the EFT coefficients
using  the simplified 9-parameter framework described in  Section 3.  
The final column gives the result of the full EFT fit described in
Section 4.   All of these fits assume data samples of 2 ab$^{-1}$ at
250~GeV.  The effective couplings 
$g(hWW)$ and $g(hZZ)$ are defined as proportional to the square root of
the corresponding partial widths.   The last two lines give 95\%
confidence upper limits on the exotic branching ratios.}
\label{tab:simplefits}
\end{center}
\end{table}

The full details of our treatment of the $\ee\to Zh$ cross section and
angular distribution are discussed in \cite{BFJPT}.   Our approach can
be summarized by saying that the perturbations to the SM cross section
from the $\eta_Z$ and $\zeta_Z$ terms can be completely 
parametrized in terms of
coefficients $a$ and $b$, respectively, that describe the variations
in the cross section, angular distribution, and $Z$ polarization.
These $a$ and $b$ parameters depend on beam polarization and center of
mass energy.  For example, at the tree level, the total cross section
for $\ee\to Zh$ from a  fully polarized $e^-_Le^+_R$ or $e^-_Re^+_L$
initial state is given by 
\beq
\sigma = {2\over 3} {\pi \alpha_w^2\over c_w^4} {m_Z^2\over (s -
  m_Z^2)} {2 k_Z\over \sqrt{s}} (2 + {E_Z^2\over m_Z^2} )
\cdot\,   Q_Z^2 \cdot \biggl[ 1 + 2a  + 2\ { 3 \sqrt{s}    
   E_Z/m_Z^2\over   (2 + E_Z^2 / m_Z^2)}\  b\biggr]
\eeq{genZhform}
where $k_Z$, $E_Z$ are the lab frame momentum and energy of the $Z$. 
In the simplified parameter set used here, the parameters in
\leqn{genZhform} are given, for a fully polarized $e^-_Le^+_R$ initial
state, by
\beqa
 Q_{ZL} &=& (\half-s_w^2) \ , \qquad   a_L = -c_H/2 \CR
b_L &= &  c_w^2 \bigl( 1 + {s_w^2\over 1/2 - s_w^2}{s - m_Z^2\over
  s}\bigr) (8c_{WW}) 
\eeqan
and,  for a fully polarized $e^-_Re^+_L$ initial state, by
\beqa
 Q_{ZR} &=& (-s_w^2) \ , \qquad   a_R = -c_H/2 \CR
b_R & =  & c_w^2 \bigl( 1 -{s - m_Z^2\over
  s}\bigr)  (8c_{WW}) \ . 
\eeqan
The change of sign between the two terms in $b_L$ vs. $b_R$ is the
polarization effect described earlier in this section. There are
similar
formula for the distributions in production angle and $Z$ decay
angles; see \cite{BFJPT} for details.  Fits
for the $a$ and $b$ parameters using ILD full simulation data are 
described in \cite{Ogawa}, and these are the basis for the error
estimates and correlations for these parameters listed in the Appendix.

The results of the simplified  fits are shown in
Table~\ref{tab:simplefits}. 
  We assume 2000~fb$^{-1}$ of data, equally divided
between
the two polarized beam configurations.  These fits include only data
from $\ee\to Zh$. 
The $\kappa$ fit is limited by
the poor knowledge of the Higgs total width.   In this fit, the width
is obtained through the relation \leqn{usualGamma} and suffers from a 
lack of statistics for the $h\to ZZ^*$ decay.   In the EFT fits,
the uncertainty in the couplings reflects the uncertainty in the
knowledge
of the $\zeta_Z$ parameter, which is determined mainly  from the 
data on the  reactions $\ee\to Zh$. 
 Note that polarization is in general a more powerful 
analyzer for $\zeta_Z$ than the angular distributions,
although either method can be effective with a sufficiently large
 luminosity sample.   For
reference, the last column of the fit gives the results of the full
EFT fit described in the next section.  The simplified  fit is quite
idealized,  but its outcome turns out to be close to that of a full
EFT  analysis.

\section{Projections for ILC at 250 GeV}

A complete analysis within the EFT framework requires a much larger
number of parameters.  As stated above, there are 84 possible
dimension-6 operators with the gauge symmetry and particle content of
the Standard Model.   In \cite{BFJPT}, we point to a subset of 9 of
these operators that contribute to processes involving $W$, $Z$,
Higgs, and light leptons only.  Five additional operators modify the
Higgs
couplings to fermions and gluons.  Two further parameters are needed
to describe the
$W$ and $Z$ couplings to quarks.
  Then our model-independent fit will involve 16
dimension-6 operator coefficient plus the 4 relevant parameters of the
Standard Model and the 2 parameters introduced above for exotic
Higgs decays.   The total number of parameters is   22. 
Though many papers have been written 
about fits to Higgs data using EFT, it seems not to have been realized 
that data from future $\ee$ colliders will  completely constraint
these 22 parameters,  allowing precise analyses that are
model-independent
to the extent that this subset of operators gives a general
description of new physics.

The  subset of 9 operators noted above 
 does not include the most general operators
with
$W$, $Z$, Higgs, and light leptons.   It excludes 4-fermion contact
interactions, which, however,  do not
contribute to the observables of relevance here. It 
 assumes muon-electron universality, which can be strongly tested at
 $\ee$ colliders  in
 $W$ decays and in 2-fermion scattering processes.  It  excludes
 CP violating operators.   But CP violating operators contribute only
 in order $c_I^2$  to the observables we consider, and there are other
 observables linear in these $c_I$ (for example, the forward-backward
 asymmetry in $\ee\to Zh$)  that can bound them at the percent level.  It
 excludes the coefficient $c_6$ that shifts the triple-Higgs coupling,
 which does not contribute at tree level to the observables we
 consider here~\cite{detsix}.   Some additional qualifications are
 given Section 2.2 of \cite{BFJPT}.
Most importantly, our analysis assumes that operators of dimension 8 are
 negligible and operators of dimension 6 can be treated in linear
 order only.  This makes sense for operators that provide few-percent
 corrections to the Higgs couplings, corresponding to the sensitivity
 of future $\ee$ colliders. In this paper, we will treat the dimension
 6 operators at the tree level only.  If  corrections to Higgs
 couplings turn out to be 
 at the 30\% level, one might question this assumption. But unless
 corrections to the Higgs couplings are very large, 
our  restricted---but still
22-dimensional---parameter set  can be 
considered a model-independent description of new physics for the
purpose of Higgs coupling analysis.

The fit parameters are the following:   First, since dimension-6
effects renormalize the parameters of the Standard Model, we must
include deviations in the 4 Standard Model parameters
\beq 
       \delta g, \delta g', \delta v, \delta \lambda 
\eeqn
In the basis chosen in \cite{BFJPT}, the 9 EFT parameters
for the Higgs and electroweak boson sector are 
\beq
   c_T,  c_{WW}, c_{WB}, c_{BB}, c_{3W}, c_{HL} , c_{HL}^\prime,
   c_{HE}, c_H  \ . 
\eeqn
Of these, the parameters $c_{T}$ is essentially the $T$ parameter of
precision electroweak analysis.  The parameters $c_{HL}$, $c_{HL}^\prime$,
$c_{HE}$ parametrize current-current interactions between the Higgs
boson and the leptons.
The parameters $c_{WW}$, $c_{WB}$, and $c_{BB}$ parametrize operators
quadratic in vector boson field strengths, and $c_{3W}$ parametrizes
the one possible operator cubic in the $W$ field strength.  We need 5
additional coefficients to describe the shifts in the Higgs couplings
to fermions
and gluons~\cite{gapology},
\beq
    c_{Hb}, c_{Hc}, c_{H\tau}, c_{H\mu}, c_{Hg} \ .
\eeqn
For our treatment of the Higgs decays to $WW^*$ and $ZZ^*$, we need
two further combinations of EFT coefficients, called $C_W$ and $C_Z$
in \cite{BFJPT}, which are measureable from the $W$ and $Z$ boson
total widths.
 Finally, we include
the two parameters $\delta a_{inv}$ and $\delta a_{other}$ introduced
above
\leqn{ainvis}. 

Given this parameter set, we assume a linear relation between the
parameters and observables,
\beq
        {\cal O}_i  = {\cal O}_{i,SM} + V_{iJ} c_J \ . 
\eeq{linear}
Then measurements of the ${\cal O}_i$ lead to a covariance matrix for
the $c_J$, which can then be translated into projected errors on
Higgs partial widths or Higgs couplings.

As inputs to this process, we take the following information:  First,
7 quantities very well measured in precision electroweak
studies---$\alpha$, $G_F$, $m_W$, $m_Z$, $A_\ell$, and $\Gamma(Z\to
\ell^+\ell^-)$---provide 7 strong constraints on the parameter
set. Note that we do not need to make any assumption that the
electroweak corrections are ``oblique'' in the sense of \cite{PandT}. 
For the $h\to WW^*$ and $h\to ZZ^*$ partial widths, we also must input
the measured values of the $W$ and $Z$ total widths, as described in
\cite{BFJPT}. 
At the level of dimension-6 operators, $\ee\to W^+W^-$ provides three
independent new physics parameters---$g_{1Z}$, $\kappa_A$, and
$\lambda_A$---that can be constrained by measurement.  LEP and LHC
measurements
already constrain these parameters at the 1\% level.   For this
analysis, we need more than an order of magnitude improvement in the
constraints, but we expect that this will be provided by $\ee$
colliders at the same time that they measure the Higgs boson parameters~\cite{Tevong}.
A projected covariance matrix for these parameters with 2000~fb$^{-1}$
of data at 250~GeV, estimated from ILD studies of $\ee\to W^+W^-$ at
higher energies by Marchesini~\cite{Marchesini} and Rosca~\cite{Rosca},
 is given in the Appendix.
 The LHC experiments will provide a strong measurement of the
ratio
$BR(h\to ZZ^*)/BR(h\to \gamma\gamma)$.   The ATLAS analysis
\cite{ATLASproj} estimates the error on this measurement after
3000~fb$^{-1}$ of data-taking as 3.6-4\%.   We expect that a
measurement strategy dedicated to cancelling systematic errors between
these two similar and characteristic processes can reach the
statistics limit of 2\%.   We use this latter number in our fit, but
the results are not changed if the error is indeed 4\%. This provides the
strongest
constraint on $c_{BB}$.  The expected
LHC measurement of $BR(h\to Z\gamma)/BR(h\to \gamma\gamma)$, with an
error of 31\%~\cite{ATLASZgam}, also provides a significant constraint
on $c_{WB}$.  We also input the expected LHC measurement of $BR(h\to
\mu^+\mu^-)/BR(h\to \gamma\gamma)$~\cite{ATLASproj}.  The input 
values used are listed in the Appendix.

At this stage, the measurements described constrain
 13 of the original 22 parameters.   This includes all
but 2 of the 4+9 parameters relevant to $W$, $Z$, Higgs, and light
lepton
processes.   The remaining two free parameters are $c_H$ and
$c_{WW}$, the parameters of the simplified fit described  in the previous
section.   These parameters are constrained by meausurements of the
total cross section, angular distribution, and polarization
asymmetries    in $\ee\to Zh$, as explained in the previous section.

The remaining 9 parameters are those that appear only in expressions for the  the
Higgs boson partial 
widths.  To include these parameters, we add measurements of
$\sigma\cdot BR$ for $\ee\to Zh$ followed by Higgs boson decay to
specific final states.  The decay widths to $WW^*$ and $ZZ^*$  also
depend on $c_H$ and $c_{WW}$ in a way that put additional constraints
on these parameters.   This makes the fit for these parameters more
robust and decreases its dependence on any particular input. 

Though Higgs production through $W$ fusion has a small cross section
at 250~GeV, we  include  the measurement of the rate for 
$\ee\to \nu\bar\nu h$,  $h\to b\bar b$.

The full fit contains a number of mechanisms for constraining
  the $\zeta_W$ and $\zeta_Z$ parameters, or, alternatively through
  \leqn{zetaWZ}, the EFT coefficients  $c_{WW}$, $c_{WB}$, and $c_{BB}$.
  We have explained in Section 3 how these parameters are constrained
by  measurements of $\ee\to Zh$.   The partial widths
  for Higgs decay to $WW^*$ and $ZZ^*$ also contain these parameters,
  so measurements of Higgs processes with vector bosons in the final
  state give constraints.  The coefficient $c_{WB}$ is directly
  constrained by measurements of $\ee\to W^+W^-$ angular
  distributions.   Linear combinations of the three coefficients give
  new tree-level contributions to the Higgs decay widths to
  $\gamma\gamma$ and $Z\gamma$, so that the LHC measurements of ratios
  of branching ratios provide constraints.   The effect of the various
  inputs in determining these coefficients is shown as a systematic
  progression in Table~2 of \cite{BFJPT}.    The fact that several
  different inputs contribute decreases the importance of
  beam polarization for achieving accurate determinations of the Higgs
  boson couplings with respect to the results of the simple fits
  described in Section 3.

On the other hand, there is another effect that must also be
accounted.  The EFT formalism leads to new contact interactions, of
which an example is the third diagram in
Fig.~\ref{fig:ZAdiagrams}~\cite{Shaouly}.    As noted above, these
diagrams are enhanced
by a factor $s/m_Z^2$.   They are strongly constrained only 
when the full set of Higgs
processes measurable at $\ee$ colliders is included.   The influence
of these diagrams, and the role of the inputs in controlling them, is
described in some detail in Section~5 of \cite{BFJPT}.

The results of the full 22-parameter fit are shown in the last 
column of Table~\ref{tab:simplefits}.   The results are quite similar
to those from the simple fit of the previous section.   The
introduction of many new parameters does not decrease the quality of
the fit, since the additional measurements constrain these parameters
strongly.  The largest effect is seen in the $hWW$ and $hZZ$
couplings, where a contribution from $c_{WB}$ adds a small amount to
the total error.   The full fit also depends much less strongly on
beam polarization, since this is now only one of several constraints
that determine the parameter $c_{WW}$.   We will see this in the
examples presented in the next section.

Table~\ref{tab:case} gives a comparison of this EFT fit to previous
results that we have presented in the past
 for ILC, using  the $\kappa$ framework but including measurements
at 500~GeV to sharpen the determination of the Higgs total width.
The first column shows the result of this fit.  The second and third
columns give the results quoted in \cite{ILCcase} for the initial and full
phases of the ILC program described in \cite{parameters}.   The fourth
column shows the results of the fit described in Section~6 of this paper.
It is interesting that, with the new analysis method that we present here,
a long run at 250~GeV gives considerable power
for learning about the Higgs couplings even before we go 
to higher energy.   Eventually, of course, we
must do both.   Running at 500 GeV and above is also needed to complete the 
program of precision Higgs measurements by measuring the $ht\bar t$
coupling and the triple Higgs coupling.

\begin{table}
\begin{center}
\begin{tabular} {l|cccc}
 &full  250 GeV EFT fit  & initial ILC  \cite{ILCcase} & full ILC
                                                      \cite{ILCcase}
& full ILC EFT fit
\\
\hline
$g(hb\bar b)$ &      1.04   &      1.5 &  0.7   &  0.55  \\ 
$g(h c\bar c)$ &     1.79    &  2.7   &   1.2 &  1.09\\ 
$g(h gg)$ &     1.60     & 2.3  &  1.0   &  0.89  \\ 
$g(hWW)$ &     0.65   &  0.81 &  0.42  & 0.34 \\ 
$g(h\tau\tau)$ &   1.16  & 1.9    &   0.9  & 0.71 \\ 
$g(hZZ)$ &            0.66  & 0.58 &   0.31 &  0.34 \\ 
$g(h\mu\mu)$ &    5.53  &   20 &  9.2  & 4.95  \\ 
\hline
$\Gamma_h$ &    2.38    & 3.8  &   1.8   &  1.50  \\ 
\end{tabular}
\caption{Projected relative errors for Higgs boson couplings and other Higgs
observables, in \%, from the EFT fits in this paper, compared to the
results of Higgs couplings fits shown in  Table 1 of \cite{ILCcase}.
The first column gives the result of the fit described in Section 4,
with 2 ab$^{-1}$ of data at 250 GeV.  The fourth column gives the
results of Section 6, adding 4 ab$^{-1}$ at 500~GeV.  The total data
samples assumed in the third and fourth columns are the same.  }
\label{tab:case}
\end{center}
\end{table}

\section{Polarization vs. luminosity vs. energy}

Beam polarization played an important role in the fits described in
Sections 3 and 4. However, as we pointed out in Section 4, inclusion
of the full set of observables that can be measured at $\ee$ colliders
take some pressure off the requirement for beam polarization.
Designs for the proposed circular $\ee$ colliders CEPC and
FCC-ee anticipate larger event samples than ILC at 250 GeV, but do not 
anticipate longitudinally polarized beams.  The proposed CLIC linear
collider
anticipates its initial run at a higher energy of 380~GeV.  
It is interesting to explore the trade-offs between these proposed
programs.

This question can be answered within the EFT formalism by using the 
data in the Appendix to estimate the inputs for the various
accelerator schemes, and
then performing a fits analogous to that of Section 4. 
The results  are shown in
Table~\ref{tab:polfits}.    The first column again shows the fit of
Section 4, with polarized beams and 2 ab$^{-1}$ of integrated
luminosity.    The second column shows the result for the same
integrated
luminosity at 350~GeV in the center of mass, appropriate for the 
proposed CLIC linear collider.  (The CLIC proposal 
now considers running at 380~GeV, but all published studies have been
done assuming 350~GeV.  The CLIC proposal includes additional stages
at  higher energy that are not considered here~\cite{CLIC,CLICdisclaimer}.)
 The third  and fourth columns show the results for unpolarized 
beams~\cite{Wfrompol}
using the luminosity samples of 5 ab$^{-1}$, projected for
CEPC~\cite{CEPC},
and for 5~ab$^{-1}$ at 250~GeV plus 1.5~ab$^{-1}$ at 350~GeV,
approximating the program projected for FCC-ee  with 2
detectors~\cite{FCCee}.   Our error estimates include an
accounting for expected 
systematic errors, as described in the Appendix. 

 The  fifth
column  shows the results of a fit including data at 500 GeV that will
be described in the next section.  This fit include 2 ab$^{-1}$ at
250~GeV
plus 4~ab$^{-1}$ at 500~GeV, realizing the full ILC plan set out in
\cite{parameters}.

\begin{table}
\begin{center}
\begin{tabular} {lccccc}
 &  2 ab$^{-1}$ & 2 ab$^{-1}$  & 5 ab$^{-1}$ & + 1.5 ab$^{-1}$ &  full ILC \\
 &  w. pol. &   350~GeV & no pol. &  at 350 GeV & 250+500 GeV \\ 
\hline
$g(hb\bar b)$ &        1.04  &   1.08  &  0.98     &   0.66  &  0.55 \\ 
$g(h c\bar c)$ &       1.79 &   2.27 &   1.42    &   1.15 &  1.09 \\ 
$g(h gg)$ &              1.60 & 1.65  & 1.31   &  0.99  &   0.89\\
$g(hWW)$ &              0.65 &  0.56  &   0.80   &   0.42  &   0.34 \\ 
$g(h\tau\tau)$ &      1.16  & 1.35   & 1.06   & 0.75 & 0.71 \\ 
$g(hZZ)$ &               0.66  &   0.57    &  0.80   &  0.42   &   0.34 \\ 
$g(h\gamma\gamma)$ & 1.20 & 1.15&  1.26   &  1.04   &  1.01 \\
$g(h\mu\mu)$ &       5.53  &  5.71  &  5.10 &  4.87  & 4.95  \\ 
\hline
$g(hbb)/g(hWW)$ &      0.82   & 0.90   &  0.58    &  0.51 &  0.43 \\ 
$g(hWW)/g(hZZ)$ &    0.07 &   0.06   & 0.07  & 0.06  & 0.05  \\ 
\hline
$\Gamma_h$ &           2.38 & 2.50  &     2.11   & 1.49 &   1.50\\ 
$\sigma(\ee\to Zh)$ &  0.70 &   0.77  &  0.50  &    0.22  & 0.61 \\ 
\hline
$BR(h\to inv)$ &          0.30  &  0.56 &    0.30     &    0.27  &  0.28 \\ 
$BR(h\to other)$ &       1.50  &  1.63  &  1.09     &   0.94   &  1.15
\end{tabular}
\caption{Projected relative errors for Higgs boson couplings and other Higgs
observables, in \%,  comparing the full EFT fit described in Section 4
to other possible $\ee$ collider scenarios.   The second column shows
a fit with 2 ab$^{-1}$, with 80\% electron and zero positron polarization, and with a higher energy of
350~GeV.
The third and fourth  columns show scenarios with no polarization but higher
intergrated luminosity,  5~ab$^{-1}$ at 250~GeV in the third column
and 5~ab$^{-1}$ at 250 GeV plus 1.5~ab$^{-1}$ at 350~GeV in the fourth column.
The fifth column gives  the result of the fit described in Section 6
including data from 250 and 500~GeV. The notation is as in Table~\ref{tab:simplefits}.}
\label{tab:polfits}
\end{center}
\end{table}

It is clear from the table that the decreased power of the angular
distributions to measure the $\zeta$ parameters can be compensated by
higher luminosity.   One should also remember that the ratios of
branching ratios are measured at $\ee$ colliders  without ambiguity,
and the accuracy of these measurements improves as $\sqrt{N}$.   These
ratios of branching ratios can be important in the testing of specific 
models, as we will discuss in Section 7. 

One should note that beam polarization offers some qualitative
advantages that are not captured in a table such as this.   Having
separate samples with different beam polarization essentially doubles
the number of independent observables and allows consistency tests
that would be not otherwise be available.

\begin{table}
\begin{center}
\begin{tabular} {l|cccc}
 &no pol. & 80\%/0\% & 80\%/30\% \\
\hline
$g(hb\bar b)$ &                 1.33   &   1.13 &  1.04   \\ 
$g(h c\bar c)$ &                  2.09     & 1.97  &  1.79 \\ 
$g(h gg)$ &                            1.90   &1.77 &  1.60 \\ 
$g(hWW)$ &                          0.98  &  0.68 &  0.65  \\ 
$g(h\tau\tau)$ &                    1.45  & 1.27   &   1.16  \\ 
$g(hZZ)$ &                            0.97  & 0.69 &   0.66\\ 
$g(h\gamma\gamma)$ &      1.38  &  1.22 &  1.20\\
$g(h\mu\mu)$ &                    5.67  &  5.64 &  5.53   \\ 
\hline
$g(hb\bar b)/g(hWW)$ &     0.91   &  0.91  & 0.82         \\ 
$g(hWW)/g(hZZ)$ &            0.07 &   0.07  & 0.07   \\ 
\hline
$\Gamma_h$ &                      2.93 &  2.60  & 2.38  \\ 
$\sigma(\ee\to Zh)$ &        0.78 &  0.78  & 0.70       \\ 
\hline
$BR(h\to inv)$ &                0.36 &    0.33    &   0.30      \\ 
$BR(h\to other)$ &          1.68   &      1.67 &   1.50
  \\ 
\end{tabular}
\caption{Projected relative errors for Higgs boson couplings and other Higgs
observables
with 2 ab$^{-1}$ of data at 250 GeV, comparing the cases of  zero
polarization, 80\% $e^-$ polarization and zero positron polarization,
and 80\% $e^-$ polarization and 30\% positron polarization.  In each
case, the running is equally divided into two samples with opposite
beam polarization orientation.}
\label{tab:poltest}
\end{center}
\end{table}

It is amusing to use comparisons such as this one to try to determine
the 
``best'' future $\ee$ collider, but truly the best collider is the one that
is actually built.   The trade-off shown between linear and
circular colliders shows that colliders of both types have powerful
capability for discovering new physics beyond the Standard Model.  We
will present an explicit comparison with models for ILC in Section 7.
Similar results would be obtained with any of the scenarios shown in
Table~\ref{tab:polfits}. 

It is of some interest to understand the importance of positron
polarization, in addition to electron polarization, since positron
polarization at linear colliders requires a special type of positron
source.  Table~\ref{tab:poltest} investigates this question
by comparing the results of a complete EFT fit at 250 GeV and
2000~fb$^{-1}$ for different assumptions about the electron and
positron polarization.

\section{Inclusion of $\ee$ data at 500~GeV}

Our discussion so far has focused mainly on $\ee$ data that might be
collected at 250 GeV.   The ILC envisions a stage of running at
500~GeV.  For CLIC, the current plan is to initially run at 380~GeV,
with subsequent stages at 1~TeV and above.   Thus it is interesting to
consider the effect of higher-energy data on this analysis.

There are three important effects of higher-energy running.   First, the
$W$ fusion process  $\ee\to \nu\bar\nu h$ turns on, providing a new
source of data on Higgs cross sections and branching ratios.   The
dependence of this cross section on the parameter $\zeta_W$ is rather
weak,
\beq
   \sigma( \ee\to \nu\bar\nu h )/(SM) = \cases{ (1 + 2 \eta_W -0.22
   \zeta_W ) &    $E_{CM} = 250$~GeV\cr 
(1 + 2 \eta_W - 0.34
   \zeta_W ) &    $E_{CM} = 380$~GeV\cr 
(1 + 2 \eta_W - 0.39
   \zeta_W ) &    $E_{CM} = 500$~GeV\cr }
\ .
\eeqn
Also, since the Higgs bosons from this reaction are not tagged, it is
not possible to directly measure the absolute cross section.
However, the $W$ fusion cross section is larger than the
Higgsstrahlung cross section at 500~GeV, and the luminosity of linear
colliders is expected to increase with energy, so this process adds a
large amount of information on relative Higgs branching ratios.   By
combining the cross section measurement to specific final states with
absolution branching ratio measurements at 250~GeV, one finds improved
constraints on $\eta_W$ or $c_H$. 

Second, as the center of mass energy increases, the angular
distribution of $\ee\to Zh$ predicted by the Standard Model
 evolves from one that is relatively flat to a $\sin^2\theta$
 distribution dominated by production of the longitudinal $Z$
 polarization state.  Since the contribution of the $\zeta_Z$  term is
 flat in angle and roughly independent of $Z$ polarization, this
 allows a much better discrimination of the $\eta_Z$ and $\zeta_Z$
 contributions than at 250~GeV.

Third, the effects of dimension-6 operators in $\ee\to W^+W^-$
increases as $s/m_W^2$.   Thus, running at higher energy allows
stronger constraints on new physics effects in  the triple gauge boson
couplings, improving the error on the parameter $c_{WB}$ in the manner
called for at the end of Section 4.

The results of a complete EFT fit including these effects is shown in 
the last column of Table~\ref{tab:polfits}.   This fit assumes
2~ab$^{-1}$ of data at 250~GeV plus 4~ab$^{-1}$ at 500~GeV, divided
equally between $e^-_Le^+_R$ and $e^-_Re^+_L$ polarized beams.
The input measurements, with accuracies estimated by full simulation
with the ILD detector model, are given in the Appendix.

\section{Recognition of  new physics models}

We can use the formalism presented in this paper
 to give quantitative estimates of the power of $\ee$ measurements of
 the Higgs boson 
to discover and  discriminate models of new physics beyond the SM.
  From the large literature on new physics
modification of the Higgs couplings, we have chosen a selection
of models that we feel are illustrative of possible new physics
effects on the Higgs couplings.  In this section, we will compare
their predictions for Higgs couplings to the ILC error estimates  computed in
this paper.

The EFT fit presented here can be used to estimate   the significance
of the observation of Higgs coupling deviations from the SM and
discrimination of the effects of different models.  Our method makes
use  of the linear dependence of the Higgs couplings on the EFT coefficients.
Consider as observables ${\cal O}_i$ the Higgs couplings obtained from
the fits described in this paper.   Each coupling has an expansion in 
EFT coefficients of the form \leqn{linear} with coefficients $V_{iJ}$.
   Let  $\delta g_i$ be the 
Higgs couplings deviations predicted in a given model, arranged as a vector.
Let $C_{JK}$ be the covariance matrix of the variables $c_J$
determined by the fit.  Then
\beq
         (  \chi^2)  =     g^T\ \bigl[ V C V^T]^{-1}  \ g  \  
\eeqn
gives the  $\chi^2 = 2 \log$ likelihood testing the goodness of fit
for this model relative to the Standard Model.   Similarly, if $g_A$
and $g_B$ are two such vectors for models $A$ and $B$,
\beq
         (  \chi^2)_{AB}  =    ( g_A^T - g_B^T)\  \bigl[ V C V^T]^{-1}
         \ (g_A - g_B)   \  .
\eeqn
gives the $\chi^2$ for $A$ given the hypothesis $B$ or vice versa.
The significance in $\sigma$ of the deviation of a model from the SM, or of one
model from another, is roughly the square root of the $\chi^2$
computed in this way.

\begin{table}
\begin{center}
\begin{tabular} {llccccccccc}
 &Model  & $b\bar b$ &       $c\bar c$    & $gg$  &  $WW$ & $\tau\tau$ &   $ ZZ$  &
                      $\gamma\gamma$    & $\mu\mu$   \\ \hline
1&MSSM~\cite{PMSSM}  &  +4.8 &-0.8  &  - 0.8& -0.2 & +0.4 &  -0.5 &
                    +0.1   &   +0.3\\
2&Type II 2HD ~\cite{Shinyaone} &+10.1  & -0.2 &-0.2 & 0.0&+9.8 &0.0
                                                             &+0.1  &+9.8\\
3&Type X 2HD ~\cite{Shinyaone} &-0.2  &-0.2 &-0.2 &0.0&+7.8
                                                   &0.0 &0.0 &+7.8\\
4&Type Y 2HD ~\cite{Shinyaone} &+10.1  &-0.2 & -0.2  & 0.0 &-0.2 &
                                                 0.0 &0.1 &-0.2\\
5&  Composite Higgs~\cite{Contino} &-6.4 & -6.4 & -6.4& -2.1& -6.4&
                                                   -2.1& -2.1& -6.4\\
6&Little Higgs w. T-parity~\cite{Maximmodel} &
                   0.0 & 0.0  &   -6.1   &   -2.5  &  0.0  &  -2.5
                                             &  -1.5 &    0.0 \\
7&Little Higgs w. T-parity~\cite{Yuanmodel} &
                    -7.8 & -4.6  &   -3.5   &   -1.5  &  -7.8   &   -1.5
                                             &  -1.0  &-7.8 \\
8&Higgs-Radion~\cite{HRradion} &  -1.5  &  - 1.5   & +10. & -1.5 & -1.5 & 
                  -1.5 &    -1.0 & -1.5 \\
9&Higgs Singlet~\cite{Christophemodel} &  -3.5 &  -3.5 & -3.5& -3.5 &
                                                                     -3.5 & -3.5 & -3.5& -3.5 \\
\end{tabular}
\caption{Deviations from the Standard Model predictions for the Higgs
  boson couplings, in \%, for  the  set of new physics models
  described in the text.  As in Table~\ref{tab:simplefits},  the effective couplings 
$g(hWW)$ and $g(hZZ)$ are defined as proportional to the square roots of
the corresponding partial widths. }
\label{tab:models}
\end{center}
\end{table}

It will always be true that some models of new physics are observable
through Higgs coupling deviations while others are not.
 All viable models of new
physics beyond the SM exhibit ``decoupling''.    That is, as the new
particle masses are increased, the predicted deviations from the SM
in  the Higgs
couplings and in other precision observables tend to zero. It is
interesting to ask, though, whether there are models that predict
significant deviations in the Higgs couplings for parameter values  at which
the new particles are very heavy, outside the reach of the LHC.   A
systematic 
study of supersymmetric models~\cite{PMSSM,PMSSMtwo} 
 shows that there are a significant
number of such models.  In fact, Figs.  1--5 of \cite{PMSSM} show that
constraints on the Higgs couplings proble the model space in a
direction roughly orthogonal to that probed by direct particle
searches.  It makes sense, then, to open this new line of attack on
the problem of discovering physics beyond the SM.

To illustrate the range of new physics models that can be found
through studies of the Higgs couplings, we present a list of 9 models
with significant Higgs boson coupling deviations in which the new
particles present in the model are unlikely to be discoverable  at
the LHC, even in the high-luminosity era.
 These models are:
\begin{enumerate}
\item  A supersymmetric model, model \#1259073  of the PMSSM models
  described in \cite{PMSSM}.   This model has relatively 
 high colored SUSY particle masses:  
  $m(\s b) = 3.4$~TeV, $m(\s g) =
4$~TeV, but still these shift the $hb\bar b$ coupling significantly. 
 The lightest SUSY particles are 
   a Higgsino multiplet at 515 GeV.
\item  A Type II 2-Higgs-doublet model from \cite{Shinyaone}, with
  heavy Higgs bosons at 600~GeV and $\tan\beta = 7$. 
Higgs couplings are evaluated
 with 1-loop corrections as in \cite{Shinyatwo}. This model and the
 next two lie in the wedge-shaped region where the LHC has limited sensitivity.
\item  A Type X 2-Higgs-doublet model from \cite{Shinyaone}, with
  heavy Higgs bosons at 450~GeV and $\tan\beta = 6$. 
Higgs couplings are evaluated
 with 1-loop corrections as in \cite{Shinyatwo}.
\item  A Type Y 2-Higgs-doublet model from \cite{Shinyaone}, with
  heavy Higgs bosons at 600~GeV and $\tan\beta = 7$. 
Higgs couplings are evaluated
 with 1-loop corrections as in \cite{Shinyatwo}.
\item A Composite Higgs model MCHM5 with $f = 1.2$~TeV, described in 
\cite{Contino}.   The lightest new particle in this model is a
vectorlike  top
quark   partner $T$ at  1.7~TeV.   However, the single production
cross section for this particle can be very small.
\item A Little Higgs model with T-parity in the family of models
  considered
in \cite{Maximmodel}, with $f = 785$~GeV and the top quark partner $T$
at 2~TeV.   [This model is on the boundary with respect to precision electroweak.]
\item  A Little Higgs model with T-parity  described in
  \cite{Yuanmodel}, with $f = 1$~TeV and the option B for light-quark Yukawa couplings.
In this model, the top quark partner $T$ has a mass of 
2.03~TeV.
\item  A Higgs-radion mixing model described in \cite{HRradion}.
  The radion mass is taken to be 500~GeV; other relevant
  extra-dimensional states can be at multi-TeV masses.
\item  A model with a Higgs singlet is added to the Standard Model
  to allow electroweak baryogenesis and to provide a portal to the
  dark matter sector, described in \cite{Christophemodel}.   The singlet mass is
  2.8~TeV, with mixing as large as permitted by decoupling.
\end{enumerate}
 The coupling deviations in these models are listed
in Table~\ref{tab:models}.    These deviations are shown graphically
in 
Appendix B, together with the uncertainties that would result from the fit
to ILC data at 250~GeV and 500~GeV.

\begin{figure}
\begin{center}
\includegraphics[width=0.85\hsize]{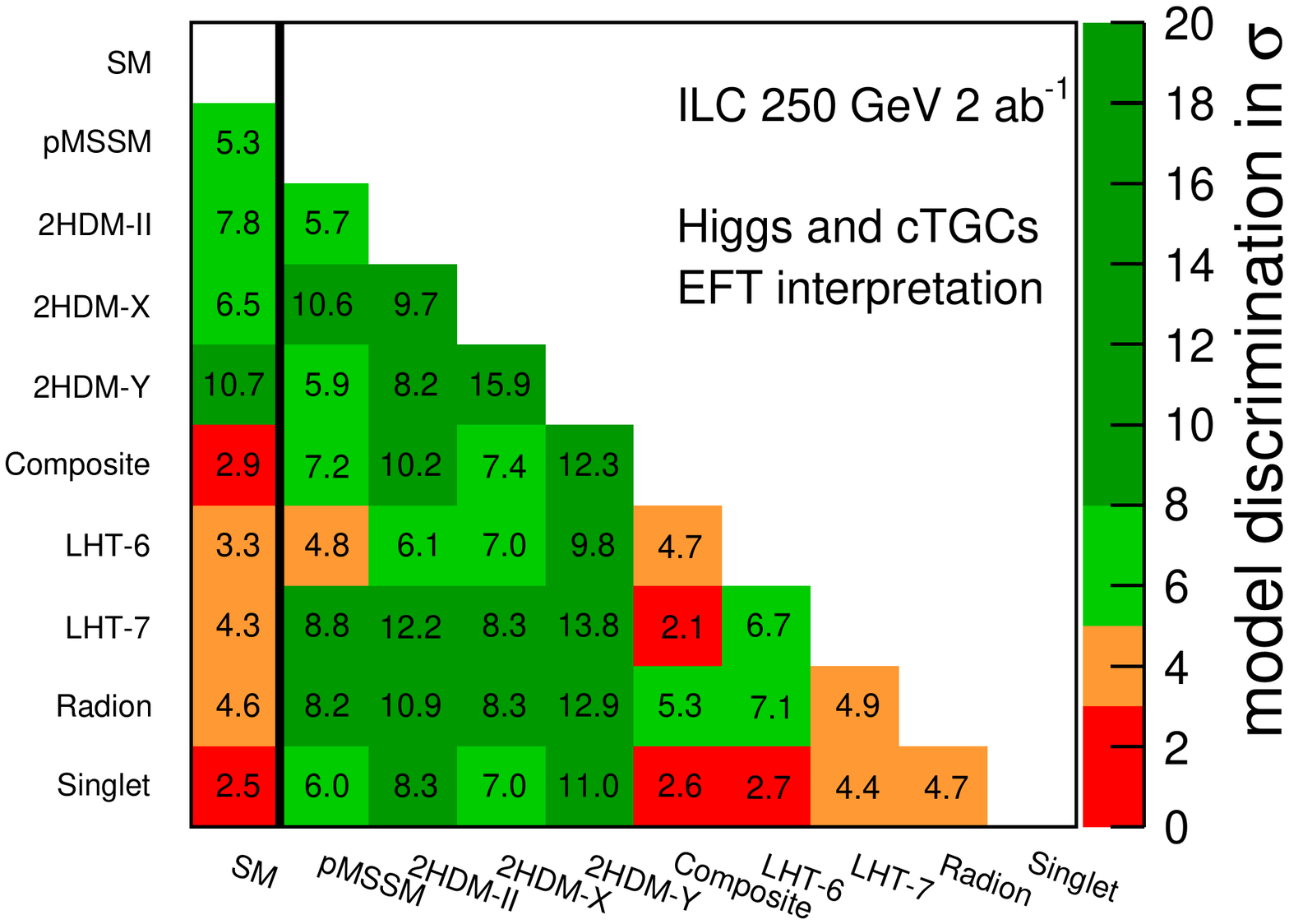} \\
\includegraphics[width=0.85\hsize]{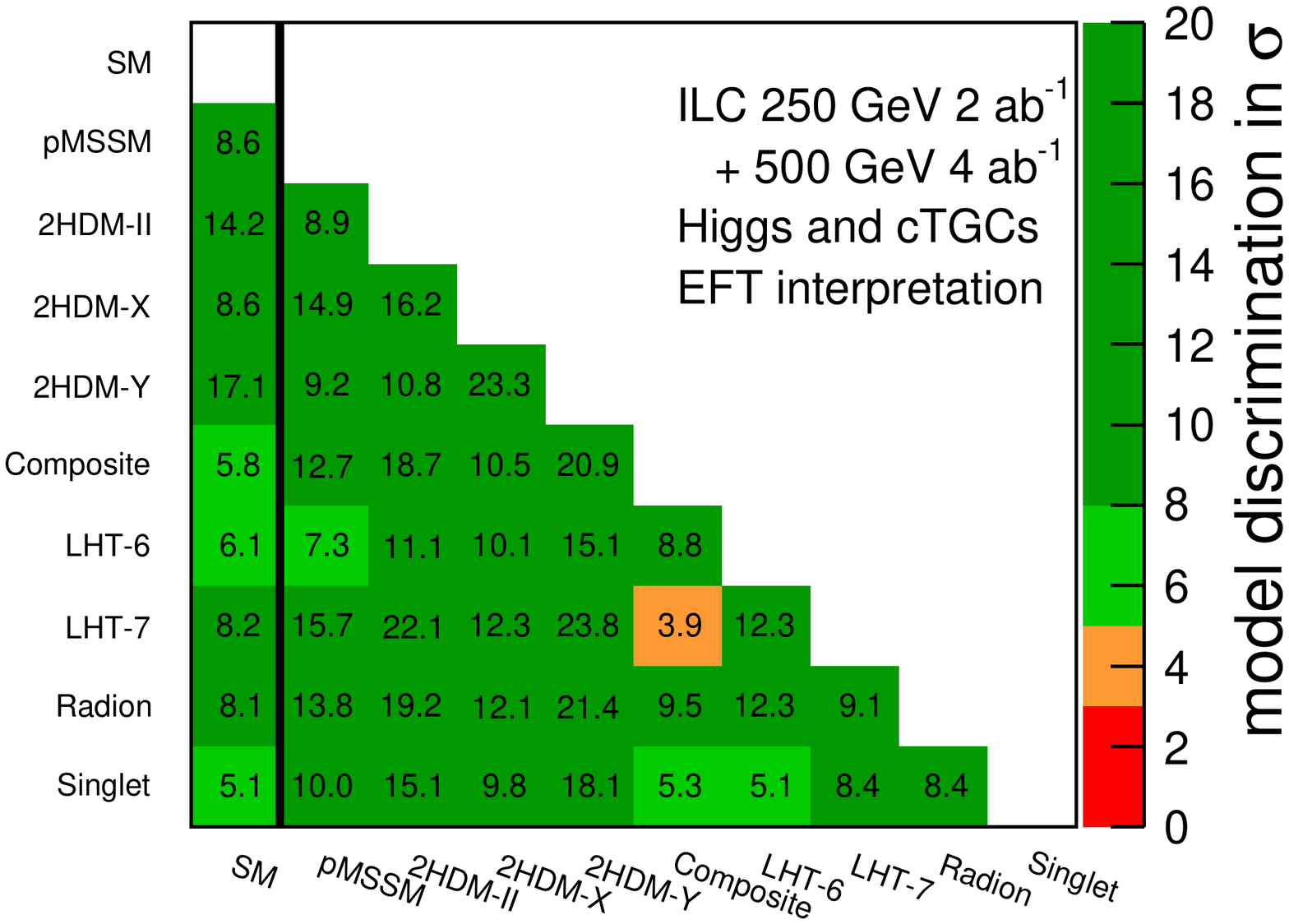}
\end{center}
\caption{Graphical representation of the $\chi^2$ separation of  the
 Standard Model and the models 1--9 described in the text:  (a) with
 2~ab$^{-1}$  of data at the ILC at 250~GeV; (b) with
 2~ab$^{-1}$  of data at the ILC at 250~GeV plus 
4~ab$^{-1}$  of data at the ILC at 500~GeV. Comparisons in orange have
above 3~$\sigma$ separation; comparison in green have above 5~$\sigma$
separation; comparisons in dark green have above 8~$\sigma$
separation.  }
\label{fig:chisq}
\end{figure}

Comparing these models to the Standard Model, we find the following 
$\chi^2$ separation. Using the fit of Section 4, for the ILC at 250
GeV
with 2 ab$^{-1}$ of data, the relative $\chi^2$s of the models are:

\begin{tabular}{l|ccccccccccc}
 &  SM    &  1& 2 &  3 &  4  & 
 5 & 6 & 7 & 8 & 9 \\ \hline
SM &   0      & 29 & 63 & 43 & 115 & 10 &    12 &  20 &    22  &  7.3  \\ 
1&  29& 0    & 34 &  113  & 36 & 53  & 24 &  78  & 68 & 37   \\ 
2    &   63 &  34 &      0  &  95&  68& 105  & 39& 149  & 120 &71\\
3    & 43  &   113  &     95 &   0 & 256& 57 & 51&  71     & 70 & 50\\
4    &   115 &36 &     68  & 256&  0 & 152&97 & 191   & 167   & 123
  \\
5  &  10 & 53& 105 & 57& 152 & 0 & 23 & 5.5 & 29 & 8.3 \\
6&  12 & 24    &   39 & 51&    97  & 23 &  0  & 46 & 51  &  8.8 \\ 
7&  20  &  78    &     149   & 71&  191 & 5.5  &  46  &0  & 26  & 21 \\ 
8& 22  & 68      &    120     &    70 &  167 & 29 &   51  & 26  &  0 &  23\\
9 & 7.3 & 37 &  71     &   50 &  123 &  8.3 &8.8 & 21  &23  &    0
\end{tabular}

\noindent Every model except \#9 is separated from the 
Standard Model by at least 
3 $\sigma$, and the models are generally separated from one another by a
comparable amount~\cite{sigmaex}.  The $\sigma$ separations of the models from the
SM and from one another are illustrated in Fig.~\ref{fig:chisq}(a). 

 Using the fit of Section 6, for the ILC at 250
GeV
with 2 ab$^{-1}$ of data and then at 500 GeV with 4 ab$^{-1}$ of data,
 the relative $\chi^2$s of the models are:

\begin{tabular}{l|ccccccccccc}
 &  SM    &  1& 2 &  3 &  4  &  5 & 6 & 7 & 8& 9\\ \hline
SM &   0  & 75  & 204 &75& 295 &35 &  39 & 68 &  67  & 27 \\ 
1& 75& 0    & 80 & 222  & 85 & 162 & 55 &  247 &  193 & 101 \\ 
2    &204  & 80& 0  & 263 & 118 & 352& 124& 492 & 371 & 230\\
3   &  75 & 222 & 263 & 0 & 543 & 113 & 104 & 152 & 147 & 97 \\ 
4    &  295 & 85 & 118 & 543 & 0 & 438 & 230 & 568 & 458& 329\\ 
5  & 35 &  162 & 352 & 113 & 438 & 0 &  78 & 17 & 93 & 30 \\ 
6 & 39 & 55 & 124 & 104 & 230 & 78 & 0 & 153 & 154 & 27\\
7&  68 & 247 &  492& 152 & 568  & 17 & 153 & 0 &  85 & 72 \\ 
8& 67   & 193 & 371 & 147 & 458 & 93 & 154 & 85 & 0 & 71\\ 
9 & 27 &101 &   230 &  97 & 329& 30 & 27 & 72 & 71 & 0 \\ 
\end{tabular}

\noindent With this data set, the various models are distinguished from the
Standard Model at least 5 $\sigma$.   Except in two cases, the models
are also well distinguished from one another, so that the results give
a clear
indication
of the nature of the new physics that has been discovered through
 the Higgs coupling
deviations.  The $\sigma$ separations of the models from the
SM and from one another are illustrated in Fig.~\ref{fig:chisq}(b). 

These examples illustrate the ability of future $\ee$ colliders to
expose  models of new physics even in cases in
which the new particles are beyond the reach of direct searches at 
 the LHC.

\section{Conclusions}

In this paper, we have presented an improved method for the extraction of
model-independent Higgs couplings from the data that will be provided
by future $\ee$ colliders.   We began by explaining that a simple
parametrization of new physics effects by rescaling the Standard Model
Higgs couplings by $\kappa_I$ parameters is not sufficiently general
to encompass the full range of models of new physics.   Instead, we
advocate  the description of new physics effects on Higgs couplings
by the coefficients of dimension-6 operators that can be added to the
Lagrangian of the Standard Model.   This Effective Field Theory
description encompasses a broader range of new physics effects. At 
the same time, it  brings in new constraints from $SU(2)\times U(1)$ gauge
invariance and from Higgs cross section measurements not previously
considered in fits for Higgs coupling.   This method draws information
from precision electroweak measurements, $\ee\to W^+W^-$, and
precision Higgs measurements, thus taking full advantage of the
richness of the information provided by future $\ee$ colliders.

Using this approach, we have analyzed the expectations of the ILC and
other proposed colliders for extracting the couplings of the Higgs
boson with percent-level accuracy and for observing and discriminating
models of new physics whose new particles are beyond the reach of LHC.
This program gives a powerful avenue to the discovery of physics
beyond the Standard Model. 

\Acknowledgements

We benefited very much from discussion of this work with many
colleagues, including Halina Abramowicz, Jim Brau, Nathaniel Craig,
Erez Etzion, Howard Haber,  Sho Iwamoto, Gabriel Lee, 
Maxim Perelstein, Philipp Roloff, Yael Shadmi, and Tomohiko Tanabe.
We are grateful to  Christophe Grojean, Ahmed Ismail, Mihoko
Nojiri, Maxim
Perelstein, and Shinya Kanemura for detailed discussions of the models 
analyzed in Section 7.  We are especially grateful to
Gauthier Durieux,  Christophe
Grojean,  Jiayin Gu, and Kechen Wang for sharing with
us insights from \cite{DGGW} that were crucial to this work, 
and for much other useful advice about  the application of 
the EFT formalism. MEP thanks Kirsten Sachs for arranging a visit to
DESY that was essential to this collaboration.  TB, SJ,  and MEP
were supported by the US Department of Energy under   
 contract DE--AC02--76SF00515.  TB was also supported by a KEK
 Short-Term
Invited Fellowship.   He thanks the KEK ILC group for hospitality
during this visit. 
  SJ was also supported by
the  National Research
Foundation of Korea under grants 2015R1A4A1042542
 and 2017R1D1A1B03030820.  KF and TO were supported
by the Japan Society for the Promotion of Science (JSPS) under
Grants-in-Aid for Science Research 16H02173 and 16H02176.  JT was
supported by  JSPS under Grant-in-Aid 15H02083.
RK and JL were  supported 
 by the Deutsche Forschungsgemeinschaft
 (DFG) through the Collaborative Research Centre SFB 676 “Particles,
 Strings and the Early Universe”, project B1.

\appendix

\section{Error estimates input into the fits presented in this paper}

The fits presented in this paper rely on error estimates for precision
electroweak, $W$ boson, and Higgs boson measurements.   In particular,
they rely heavily on uncertainties estimated for measurements that can
be carried out at the ILC at 250~GeV, 350~GeV and 500~GeV.  In this
section, we provide tables of the uncertainties we have assumed in the
fits preesents here.    These uncertainties are based on
 full-simulation studies done for the SiD and ILD detector models  at the ILC
 and CEPC.   We also specify the additional inputs from precision
electroweak and LHC measurements of Higgs branching ratios that are
used in our analysis.

Table~\ref{tab:higgserrors} gives the expected statistical errors on
Higgs cross section and branching ratio measurements for polarized
beams and for luminosity samples of 250~fb$^{-1}$.
The numbers given in Table~\ref{tab:higgserrors} are statistical errors only.  They
can be rescaled to any luminosity by dividing by the square root of
the integrated luminosity.  In our fits, we have added the statistical error of
each measurement for Higgs observables ($\sigma$ or $\sigma\cdot BR$)
in quadrature with two types of assumed common systematic
errors, $1.0\times 10^{-3}$ from theory prediction, and
$1.0\times 10^{-3}$ from luminosity and beam polarizations 
measurements \cite{Rosca}. For the $h\to b\bar{b}$ 
observables, we have also added in quadrature an additional systematic error from $b$-tagging efficiency,
taken to be 
$3.0\times 10^{-3}\times\sqrt{250/L}$ ($L$ for integrated luminosities in fb$^{-1}$) \cite{ILCWhite}.

\begin{table}
\begin{center}
\begin{tabular} {lcccccc}
-80\% $e^-$, +30\% $e^+$&  polarization: \\  \hline
 & 250 GeV   &  & 350 GeV & & 500 GeV  &  \\
 &  $Zh$ & $ \nu\bar\nu h$  &  
$ Zh$ & $\nu\bar\nu h$ & $Zh$ &
 $ \nu\bar\nu h$\\ 
\hline
$\sigma$ \cite{Recoil1,Recoil2,Recoil3,Recoil4}  &    2.0     &    &1.8  &  &  4.2   &     \\  \hline
$h\to invis.$ \cite{H2inv1,H2inv2} &  0.86   &  &1.4 &   &   3.4 &     \\
\hline
$h\to b\bar b$ \cite{H2qq1,H2qq2,H2bb1,H2bb2} &   1.3 &  8.1 & 1.5  &  1.8  & 2.5  &  0.93  \\ 
$h\to c\bar c$ \cite{H2qq1,H2qq2} &  8.3 &   &11 & 19  &   18 & 8.8 \\ 
$h\to gg$ \cite{H2qq1,H2qq2} &  7.0 &  &8.4  & 7.7 & 15  &  5.8\\
$h\to WW$ \cite{H2bb2,H2WW1,H2WW2} &  4.6 &   &5.6\ $^*$ & 5.7 \ $^*$ &  7.7   &  3.4\\
$h\to \tau\tau$ \cite{H2tautau} & 3.2 &   &4.0\ $^*$ & 16\ $^*$ &  6.1  &  9.8\\
$h\to ZZ$ \cite{ILCWhite} & 18 &   & 25\ $^*$ & 20\ $^*$ & 35\ $^*$  & 12\ $^*$    \\ 
$h\to \gamma\gamma$ \cite{H2gamgam} & 34\ $^*$ &   &39\ $^*$ &  45\ $^*$ &  47 &  27 \\ 
$h\to \mu\mu$ \cite{H2mumu1,H2mumu2} & 72\ $^*$ &   &87\ $^*$&  160\ $^*$ &  120\ $^*$ &  100\ $^*$ \\
\hline\hline
$a$ \cite{Ogawa}  &  7.6   &     &2.7 \ $^*$ &   &  4.0   &  \\
$b$ &  2.7  &    & 0.69 \ $^*$   &    & 0.70   &   \\
$\rho(a,b)$  &  -99.17 &   & -95.6  \ $^*$
                  &  & -84.8 &   \\ \hline \hline
+80\% $e^-$, -30\% $e^+$&  polarization: \\  \hline
 & 250 GeV   &  & 350 GeV & & 500 GeV  &  \\
 &  $Zh$ & $ \nu\bar\nu h$  &  
$ Zh$ & $\nu\bar\nu h$ & $Zh$ &
 $ \nu\bar\nu h$\\ 
\hline
$\sigma$   &    2.0     &    &1.8  &  &  4.2   &     \\  \hline
$h\to invis.$ &  0.61   &  &1.3 &   &   2.4 &     \\
\hline
$h\to b\bar b$  &   1.3 &  33 & 1.5  &  7.5  & 2.5  &  3.8  \\ 
$h\to c\bar c$  &  8.3 &   &11 & 79  &   18 & 36 \\ 
$h\to gg$  &  7.0 &  &8.4  & 32 & 15  &  24\\
$h\to WW$  &  4.6 &   &5.6 & 24  &  7.7   &  14\\
$h\to \tau\tau$  & 3.2 &   &4.0  & 66 &  6.1  &  40\\
$h\to ZZ$  & 18 &   & 25 & 81 & 35  & 48    \\ 
$h\to \gamma\gamma$  & 34 &   &39 &  180 &  47 &  110 \\ 
$h\to \mu\mu$  & 72 &   &87 &  670 &  120 &  420 \\
\hline\hline
$a$ &  9.1   &     &3.1  \ $^*$ &   &  4.2   &  \\
$b$ &  3.2  &    & 0.79 \ $^*$   &    & 0.75   &   \\
$\rho(a,b)$  &  -99.39 &   & -96.6  \ $^*$
                  &  & -86.5 &   \\ 
\end{tabular}
\caption{Projected statistical errors, in \%, for Higgs boson 
measurements input to our fits. The errors are 
quoted for luminosity samples of 250~fb$^{-1}$
  for $\ee$ beams with -80\% electron polarization and +30\% positron
  polarization, in the top half of the table, and with  +80\% electron polarization and -30\% positron
  polarization, in the bottom half of the table. 
  Except for the first and last segments of  each set, these are measurments
  of  $\sigma \cdot BR$, relative to the Standard Model
  expectation.
The top lines gives the error for the total cross section relative to
the Standard Model and  the 95\% confidence upper limit on the branching ratio for
Higgs to invisible decays.  The bottom lines in each half give the
expected errors on the $a$ and $b$ parameters and their correlation
(all in \%)  for $\ee\to Zh$ (see
\leqn{genZhform}.    All error estimates in this table are
based on full simulation, with the
exception of entries marked with a $^*$, which are based on
extrapolation from full simulation results. }  
\label{tab:higgserrors}
\end{center}
\end{table}

The first two lines of Table~\ref{tab:higgserrors} give estimates of 
the expected error in the total $\ee\to
Zh$ cross section and the expected 95\% confidence level upper limit on
Higgs to invisible decays, assuming the Standard Model predictions.
The following entries give the estimated errors  on $\sigma\times BR$
measurements to the given final states, using the reactions $\ee\to
Zh$ and $\ee\to \nu\bar\nu h$.  The uncertainties quoted here are for
polarized
$\ee$ beams with -80\% electron and +30\% positron polarization.  
References to the original studies are given in each line.
Most of the estimates in this table are identical to the ones
reported in the article ``ILC Operating Scenarios" \cite{parameters},
a few of them have been updated since then by new full simulation
studies and are briefly described in the following. 
The estimates for $\sigma_{Zh}\cdot BR_{WW}$ are improved by
a factor of 1.4 at $\sqrt{s}=250$ GeV,
after adding the contributions from $Z\to ll,WW^*\to l\nu l\nu/l\nu2q/4q$ 
channels \cite{H2WW2}, 
and by a factor of 1.7 at $\sqrt{s}=500$ GeV, 
after adding the contributions from $Z\to qq, WW^*\to 4q$ channels \cite{H2WW3}.
The estimates for $\sigma_{Zh}\cdot BR_{bb}$ and $\sigma_{\nu\nu h}\cdot BR_{bb}$
at $\sqrt{s}=250$ GeV are updated based on new analysis performed using 
ILD DBD simulation and reconstruction tools \cite{H2bb1}, 
and, more importantly, the correlation between them, which is $-34\%$, 
is now incorporated into the fit. 
The estimates for $\sigma_{\nu\nu h}\cdot BR_{bb/cc/gg}$ at $\sqrt{s}=350$ GeV
are updated based on new full simulation results in \cite{H2qq2}.
The estimates for $\sigma_{Zh/\nu\nu h}\cdot BR_{\tau\tau}$ are
updated to the published results \cite{H2tautau}.
The up-to-date references for all of  the estimates are indicated in the
table.

\begin{table}
\begin{center}
\begin{tabular} {lcccccc}
 & 250 GeV   &  & 350 GeV & & 500 GeV  &  \\
 &  $ W^+ W^-$  &  &  $ W^+ W^-$  &  &  $ W^+ W^-$
                                       &  
 \\\hline
$g_{1Z}$ & 0 .062 \ $^*$&   &0.033 \ $^*$ &  &0.025  &  \\
$\kappa_A $ & 0.096  \ $^*$ &   & 0.049 \ $^*$  &   & 0.034 &  \\
$\lambda_A$ &   0.077 \ $^*$ &   &0.047  \ $^*$  &  &0.037  &  \\
$\rho(g_{1Z},\kappa_A)$ & 63.4  \ $^*$ &   &63.4  \ $^*$  &  & 63.4 &  \\
$\rho(g_{1Z},\lambda_A)$ &  47.7\ $^*$ &   & 47.7 \ $^*$  &
  &47.7  &  \\
$\rho(\kappa_A,\lambda_A)$ & 35.4  \ $^*$  &   & 35.4  \ $^*$ 
 &  & 35.4  &  
\end{tabular}
\caption{Projected statistical errors, in \%, for $\ee\to W^+W^-$
measurements input to our fits. The errors are 
quoted for luminosity samples of 500~fb$^{-1}$   divided equally
between beams with 
 -80\% electron polarization and +30\% positron
  polarization and  brams with  +80\% electron polarization and -30\% positron
  polarization. The last three lines give the correlation
  coefficients, also in \%.  All error estimates in this table are
based on full simulation using the ILD or SiD detector model, with the
exception of entries marked with a $^*$, which are based on
extrapolation from full simulation results. }
\label{tab:WWerrors}
\end{center}
\end{table}

For the $\ee$ beams with +80\% electron and -30\% positron polarizations,
we generally assume that the expected errors are 
independent of the
polarization state, with a slightly higher cross section for
$e^-_Le^+_R$ being compensated by slightly lower backgrounds for 
$e^-_Re^+_L$.  Dedicated studies for +80\% electron and -30\% positron
polarization were done for the $h\to invisible$ and $a,b$
measurements, and here we quote the analysis results directly.
The $WW$ fusion process $\ee\to \nu\bar\nu h$ requires the
$e^-_Le^+_R$ initial state, and so the rate is much
smaller in this polarization configuration.  The errors quoted in the
table
are obtained by multiplying the corresponding errors in the top half
of the table by 4.1, the inverse of the square root of the relative
$e^-_Le^+_R$ luminosity.

The errors on the total cross section and angular distribution in
$\ee\to Zh$ are usefully quoted as errors on the $a$ and $b$
parameters
defined in \leqn{genZhform} and in \cite{BFJPT}.  The expected errors 
on $a$ and $b$ and their correlation $\rho(a,b)$ are shown in 
Table~\ref{tab:higgserrors}.   For $\sqrt{s}=250$~GeV and
500~GeV, these errors have been estimated in \cite{Ogawa} from ILD 
full-simulation studies for both polarization configurations.
 For $\sqrt{s}=350$~GeV, we have estimated these errors
by interpolation. 

The expected errors on the anomalous TGC coupling parameters of 
$\ee\to W^+W^-$ are shown in
Table~\ref{tab:WWerrors}. These errors are based on
extrapolation from the studies of \cite{Marchesini,Rosca}, taking into account
the dependences on $\sqrt{s}$, statistics and detector acceptances~\cite{TGCKarl}.
Further more, since the studies in \cite{Marchesini,Rosca} are performed using 
a binned fit for 3 angular distributions in semi-leptonic channel,
an additional scaling factor, 1.6-2, 
is applied to the extrapolations, to take into account the potential improvement
from an unbinned fit for 5 angular distributions in both semi-leptonic and full hadronic
channels. 
The errors for TGCs are quoted for samples of 500 fb$^{-1}$ of data, divided
equally between the two polarization states. For simplicity, we use the same
estimates for the errors at unpolarized colliders. 
In our analysis, we have added to these statistical errors the systematic errors $0.3\times 10^{-3}$
for both $g_{1Z}$ and $\kappa_A$, and $0.2\times 10^{-3}$ for
$\lambda_A$~\cite{TGCKarl}.

\begin{table}
\begin{center}
\begin{tabular} {lcccc}
Observable  &    current value &   current $\sigma$ &  future
$\sigma$ &  
 SM best fit value \\ \hline
$\alpha^{-1}(m_Z^2) $&        128.9220            &  0.0178       &  
&     (same)             \\ 
$G_F$   ($10^{-10}$ GeV$^{-2}$)    &     1166378.7        & 0.6        &
&       (same)                           \\
$m_W$  (MeV) &          80385                   &  15   &   5    &
80361  \\
$m_Z$  (MeV)  &          91187.6                  &   2.1   & &
91.1880 \\ 
$m_h$   (MeV) &      125090    &   240    &   15    &   125110   \\ 
$A_\ell $ &         0.14696      &      0.0013                   &
&          0.147937  \\          
$\Gamma_\ell$  (MeV)   &         83.984                   &   0.086
&  
&        83.995                            \\
$\Gamma_Z$ (MeV)  &         2495.2                   &   2.3   &
&              2.4943                      \\
$\Gamma_W$  (MeV) &        2085                     &    42     &
2  &     2.0888 \\
\end{tabular}
\end{center}
\caption{Values and uncertainties for precision electroweak
  observables used in this paper.   The values are taken from
  \cite{RPPEW}, except for the averaged value of $A_\ell$, which
  corresponds to the averaged value of 
  $\sin^2\theta_{eff}$ in  \cite{LEPEWWG}.  The best
  fit values are those of the fit in \cite{RPPEW}.  For the
  purpose of fitting Higgs boson couplings as described in Section 7,
  we use improvements in some of the errors expected from
  LHC~\cite{SnowmassEW} and ILC~\cite{ILCcase}.  The improved
  estimate of the $W$ width is obtained from  $\Gamma_W = \Gamma(W\to
  \ell \nu)/BR(W\to \ell \nu)$. }
\label{tab:PEW}
\end{table}

The precision electroweak inputs to our fit are shown in
Table~\ref{tab:PEW}.  For most of the entries, we have assumed the
current uncertainties, from the Particle Data Group 
compilation~\cite{RPPEW}.    For three of the values, we have assumed
improvements in uncertainties:  for the $W$ mass, from
LHC~\cite{SnowmassEW}, for the Higgs boson mass, from
ILC~\cite{Recoil1}, and, for the $W$ width, from $\Gamma_W =
\Gamma(W\to \ell \nu)/BR(W\to \ell \nu)$, using the theoretical value
of $\Gamma(W\to \ell\nu)$ from our fit and the value of $BR(W\to
\ell\nu$ that will be measured at the ILC at 250 GeV with $10^7$
$W$ pair events.

We have input the following errors on ratios of branching ratios from
the LHC, as described in Section 3:
\beqa
   \delta (  BR(h\to ZZ^*)/BR(h\to \gamma\gamma) )  & = &   2\% \CR
   \delta (  BR(h\to Z\gamma)/BR(h\to \gamma\gamma) )  & = &  31\% \CR
   \delta (  BR(h\to \mu^+\mu^-)/BR(h\to \gamma\gamma) )  & = &   12\%
   \ .
\eeqan

The full set of linear relations used in this fit, and the final $22\times
22$ covariance matrices for the fit parameters given by  the ILC 250
fit and the full ILC fit are given in files  {\tt
 CandV250.txt} and {\tt CandV500.txt}  included with the arXiv posting of this paper.

Though we used existing experimental results for the fit, it is worth emphasizing 
that the fit can benefit from several additional important
measurements for which full simulation studies 
are not yet complete. First, we plan to improve the measurement of 
$\sigma_{Zh}\cdot BR_{WW}$ at 250 GeV by including contribution from 
full hadronic channels. Second, we plan to improve the constraints on 
$h\gamma Z$ couplings by adding measurements of the branching ratio for
$h\to\gamma Z)$ 
and the cross section for  $e^+e^-\to\gamma h$.  Third, we plan  to
improve the 
constraints on the EFT coefficients  $c_{HL},c_{HL}^\prime,c_{HE}$,
by measuring the  cross sections of $e^+e^-\to\gamma Z$ for both left and right handed 
beams.  Finally, we plan to carry out a full analysis of the
meausreement of the branching ratio for $W\to e\nu$  which should
improve on the estimate given in Table~\ref{tab:PEW}. 

\section{Illustration of the discrimination of models by precision
  Higgs measurements}

\begin{figure}
\begin{center}
 \includegraphics[width=0.44\hsize]{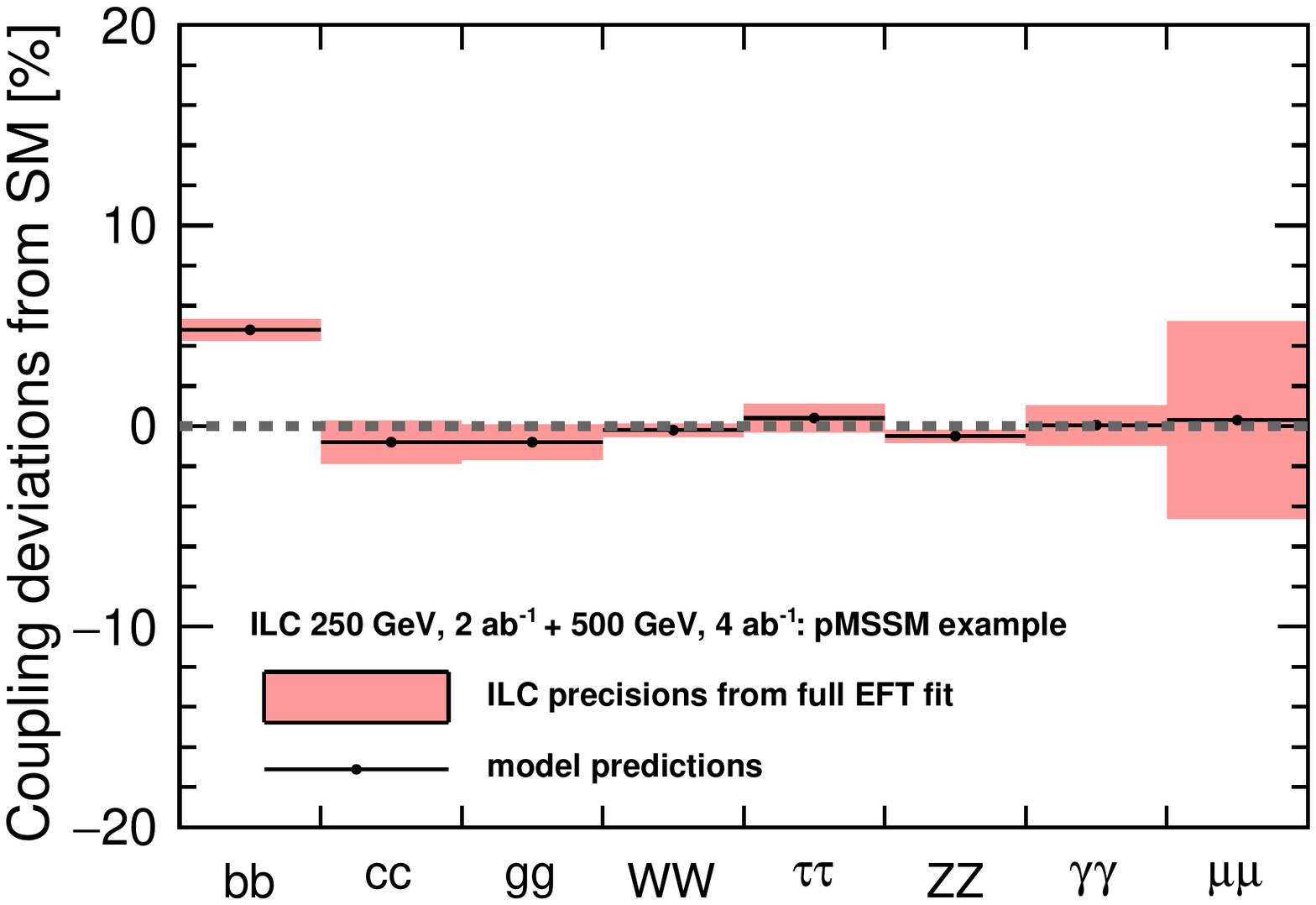} 
 \includegraphics[width=0.44\hsize]{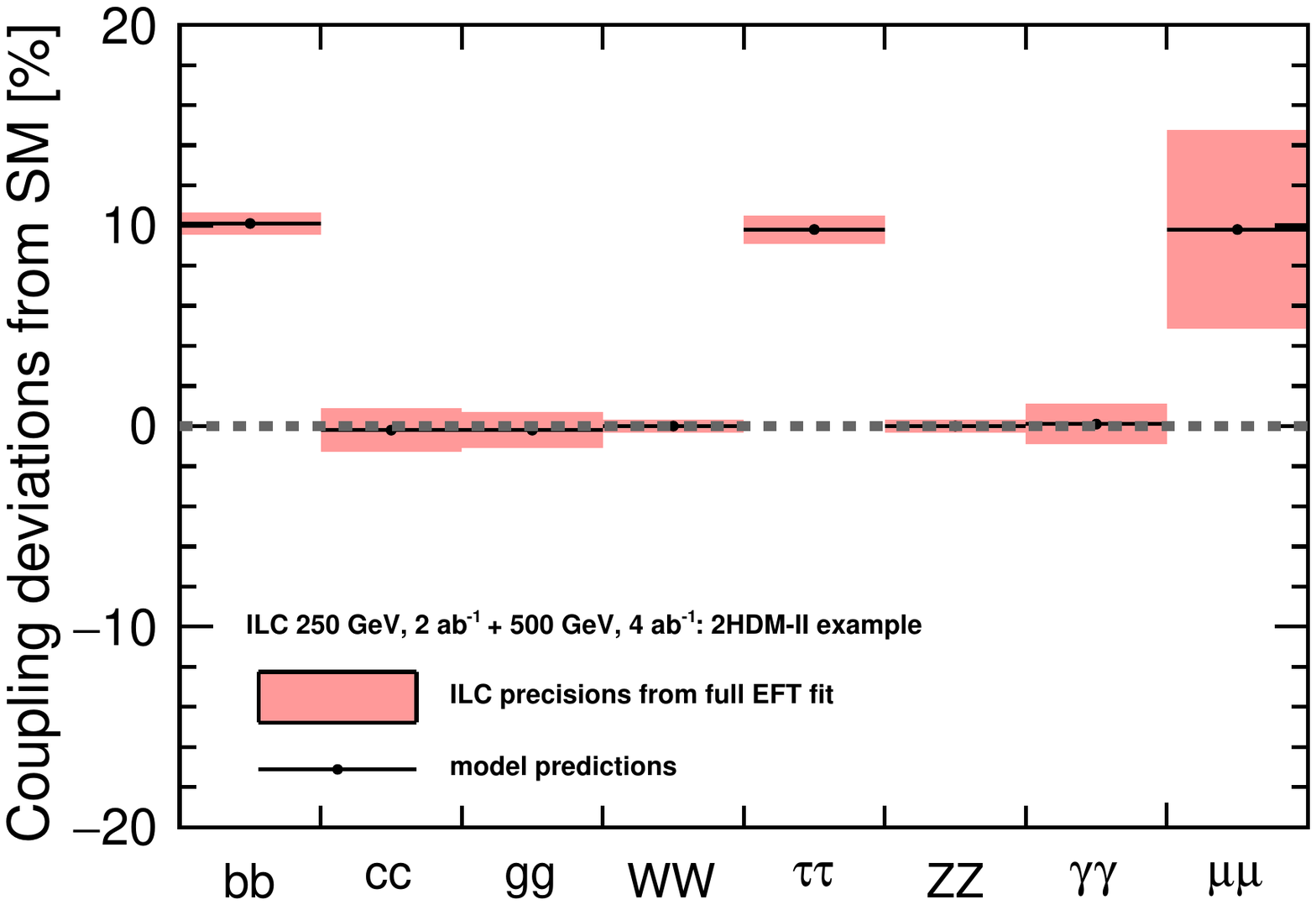}\\ 
 \includegraphics[width=0.44\hsize]{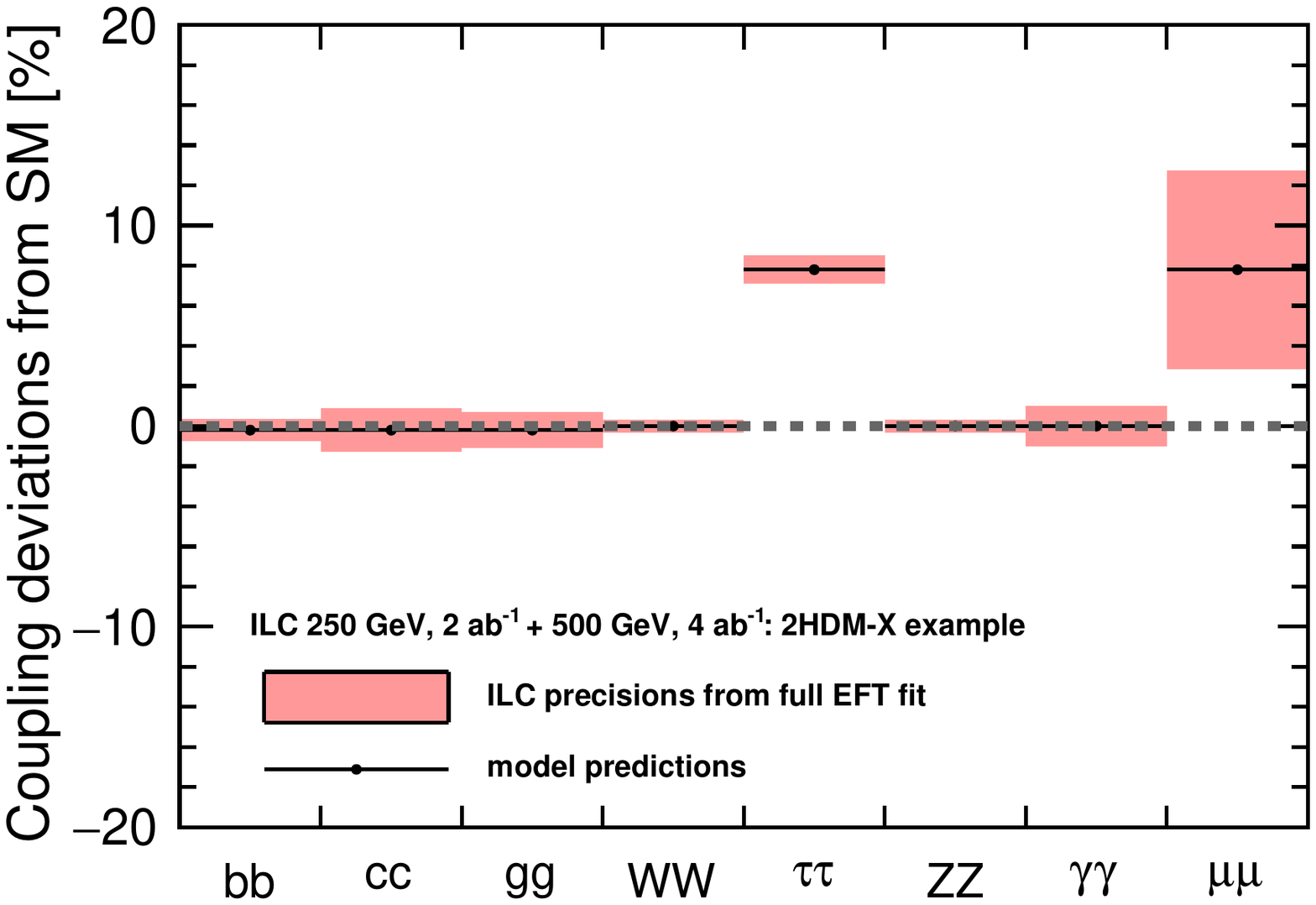} 
\includegraphics[width=0.44\hsize]{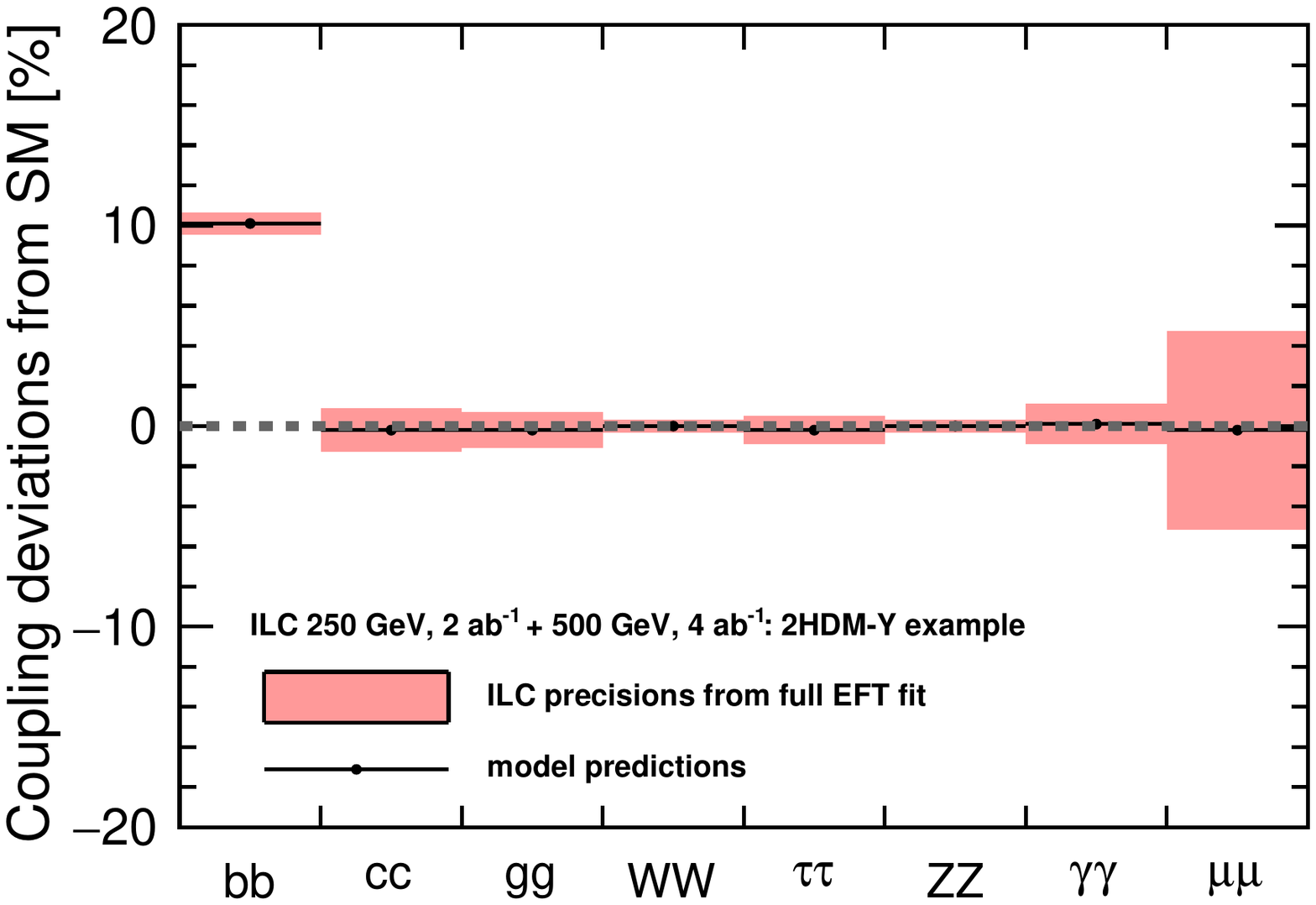}\\ 
 \includegraphics[width=0.44\hsize]{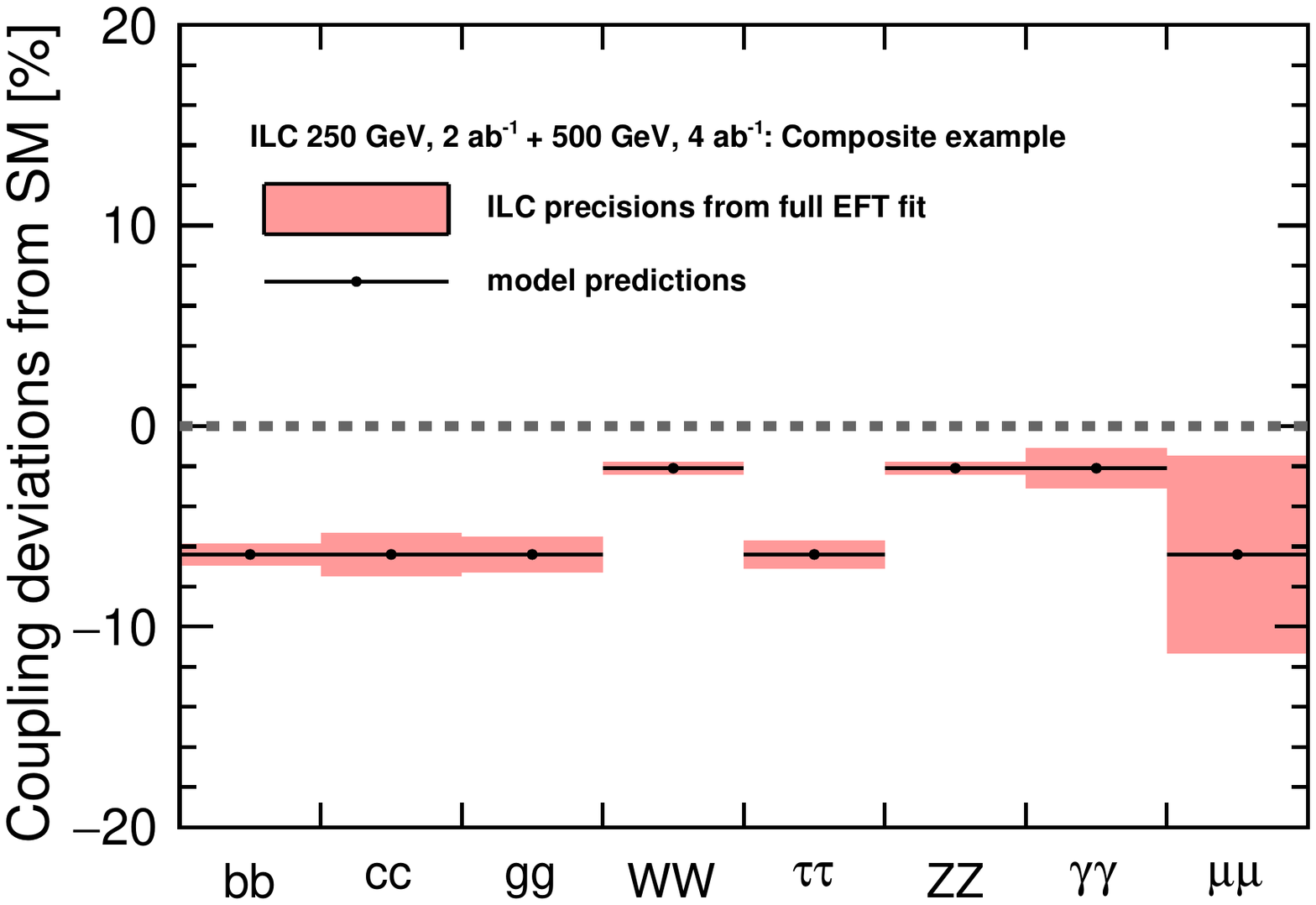}  
\includegraphics[width=0.44\hsize]{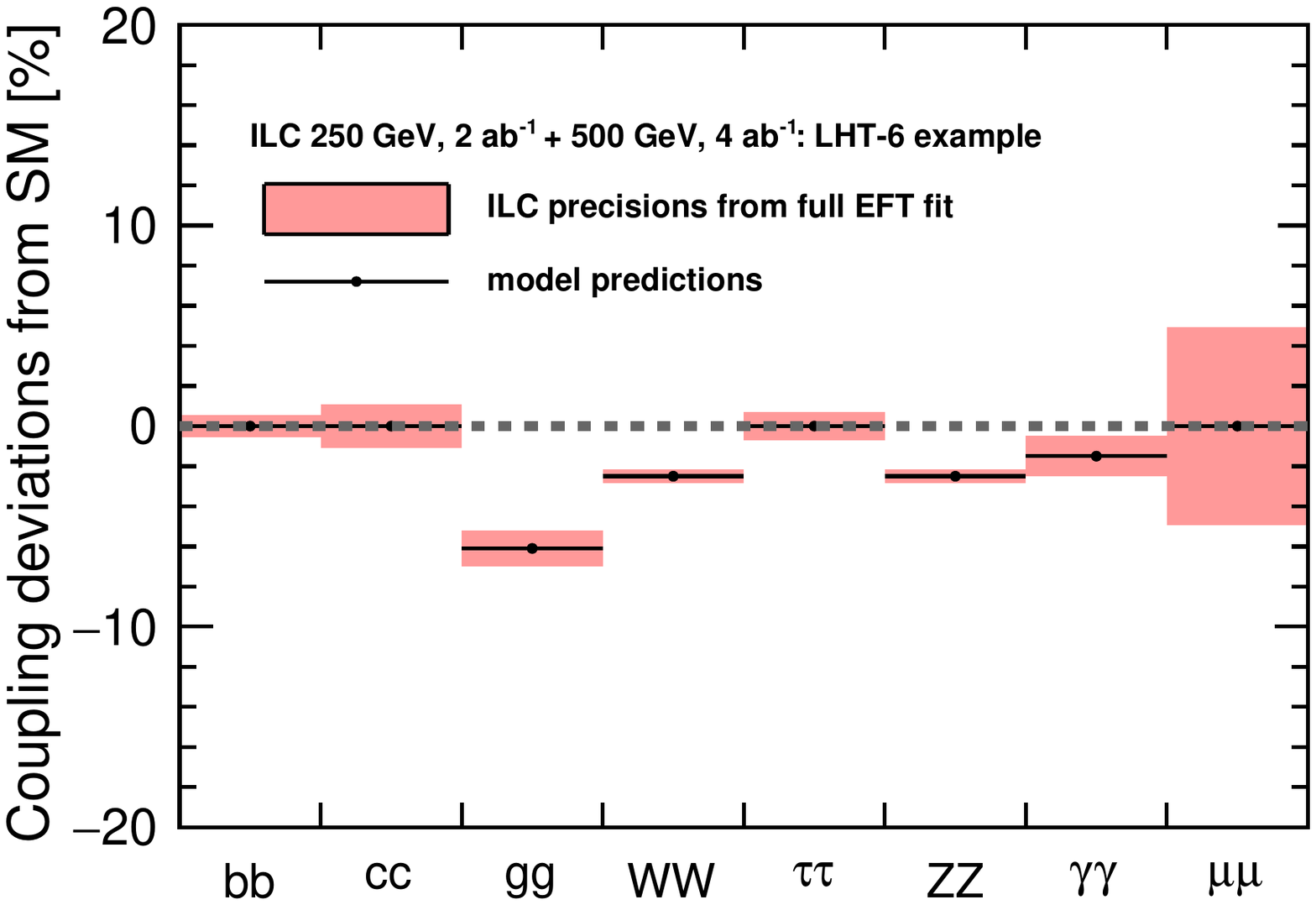}\\ 
\end{center}
\caption{Visualization of the deviations of Higgs couplings from the
  SM for the new physics models 1--6 discussed in Section 7,
  compared to  the uncertainties in the measurements expected from a fit to
  ILC data at 250 and 500~GeV.}
\label{fig:modeldis}
\end{figure}

\begin{figure}
\begin{center}
\includegraphics[width=0.44\hsize]{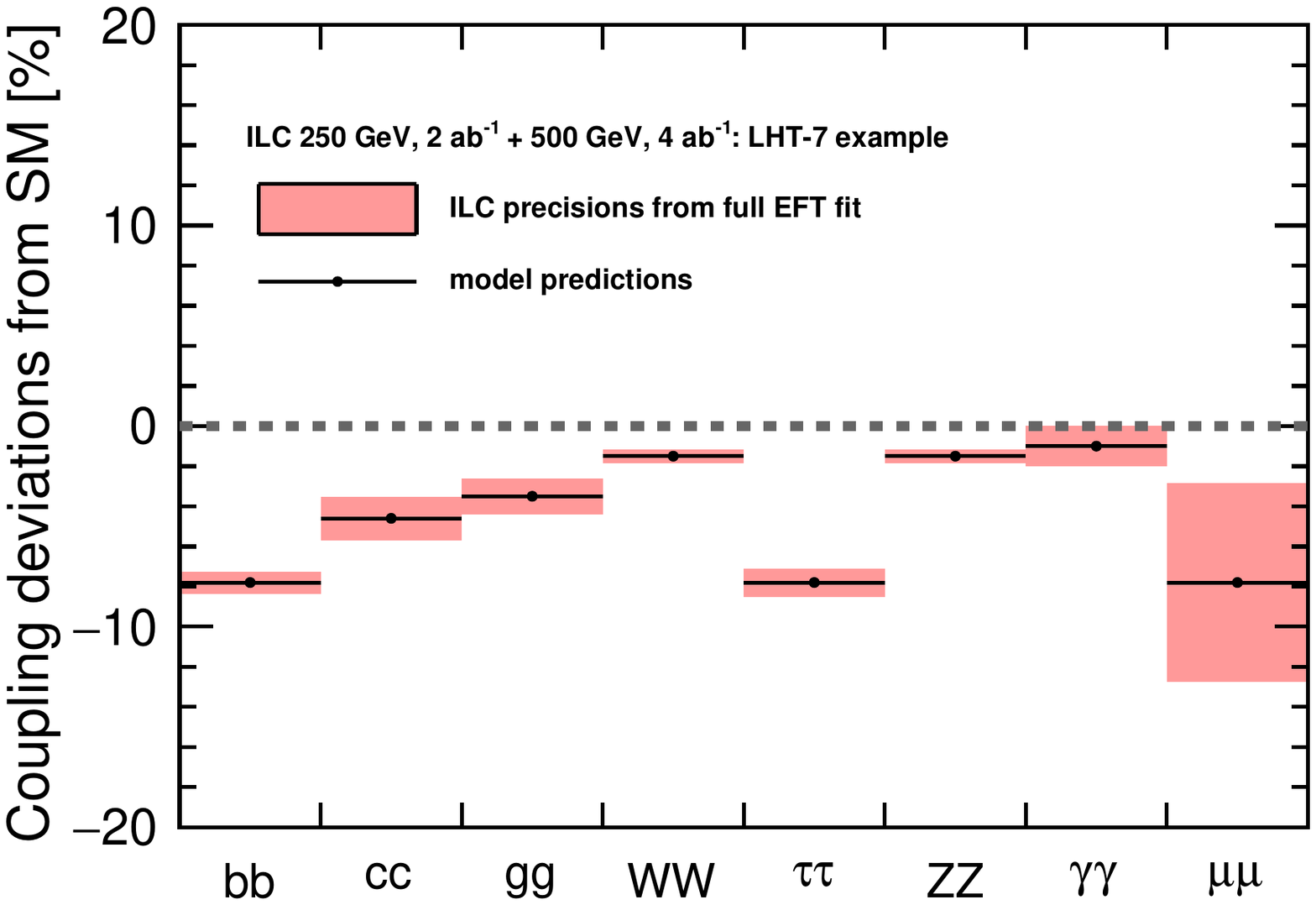} 
\includegraphics[width=0.44\hsize]{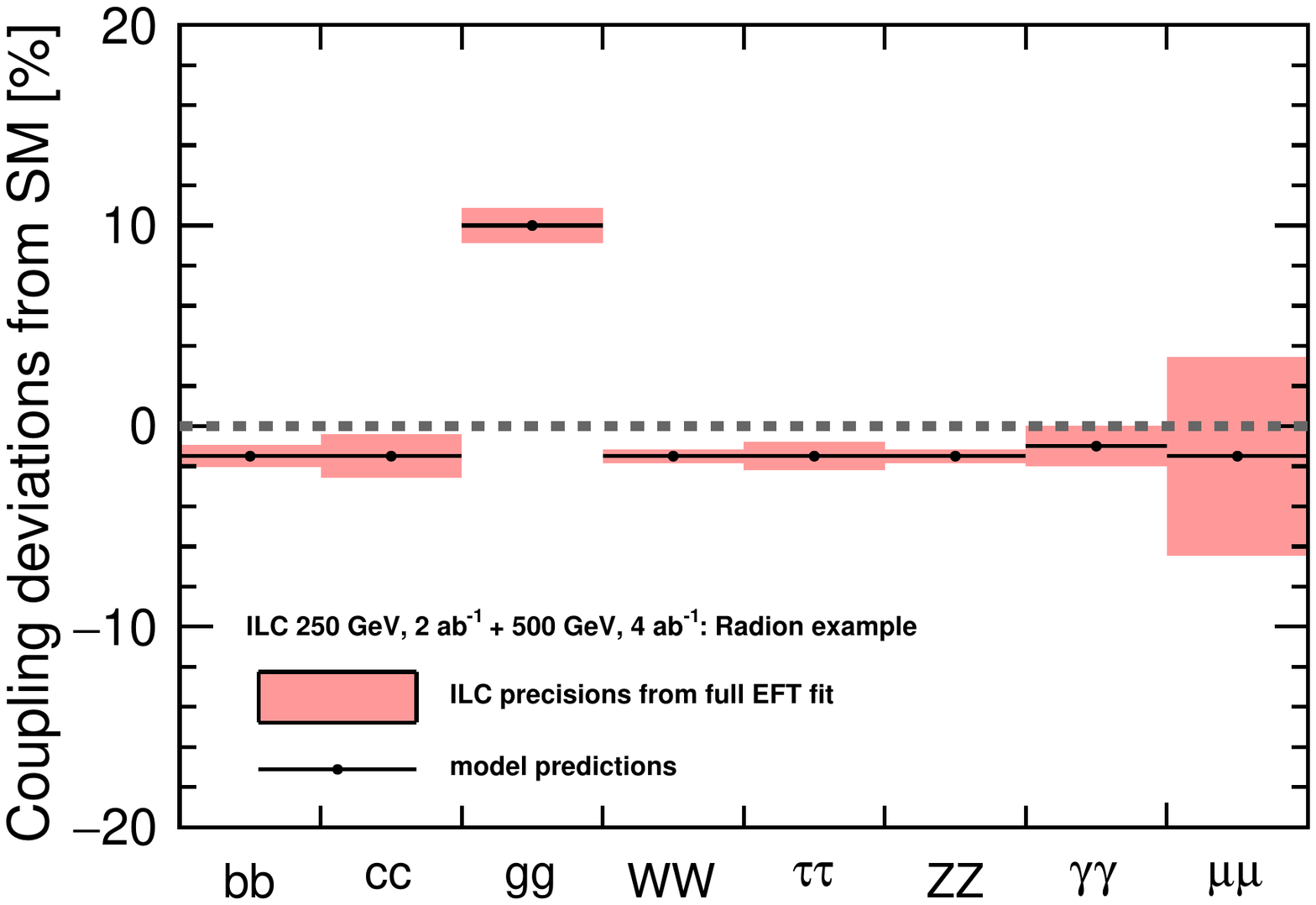}\\ 
\includegraphics[width=0.44\hsize]{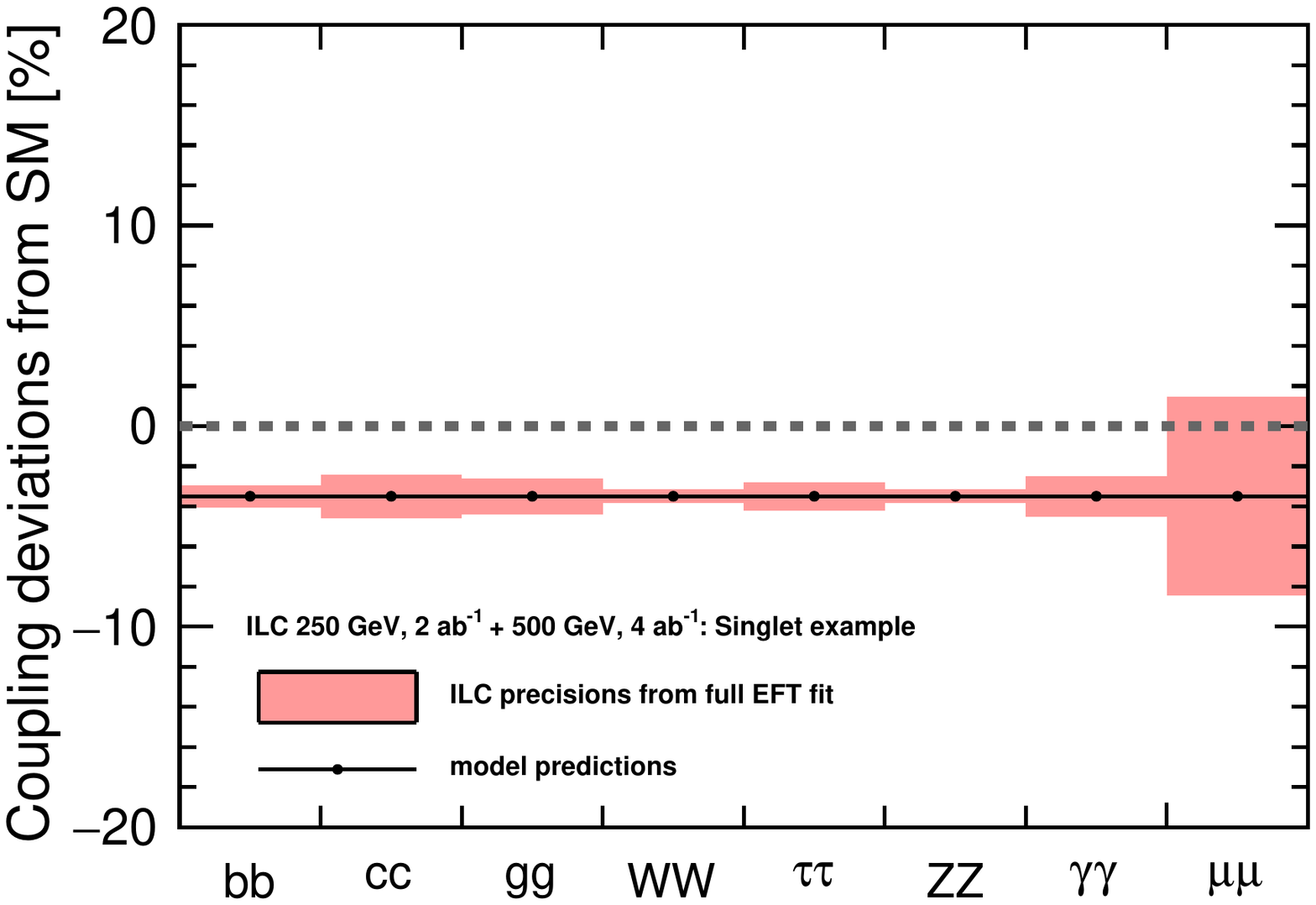} 
\end{center}
\caption{Visualization of the deviations of Higgs couplings from the
  SM for the new physics models 7--9 discussed in Section 7,
  compared to the uncertainties in the measurements expected from a fit to
  ILC data at 250 and 500~GeV.}
\label{fig:modeldisb}
\end{figure}

Figures~\ref{fig:modeldis} and \ref{fig:modeldisb} show the deviations from the Standard Model
expectations for the Higgs boson couplings, in \%,  expected for each
of the models considered in Section 7, along with the uncertainties
that would result from the fit to ILC data at 250 and 500~GeV
described in Section 6.   Note that these uncertainties  are correlated, and
that these correlations are taken into account in the $\sigma$ values
that we quote in Section~7.


\end{document}

%% file: mymacros.tex
\def\beq{\begin{equation}}
\def\eeq#1{\label{#1}\end{equation}}
\def\eeqn{\end{equation}}

\newenvironment{Eqnarray}%
   {\arraycolsep 0.14em\begin{eqnarray}}{\end{eqnarray}}
\def\beqa{\begin{Eqnarray}}
\def\eeqa#1{\label{#1}\end{Eqnarray}}
\def\eeqan{\end{Eqnarray}}
\def\CR{\nonumber \\ }

\def\leqn#1{(\ref{#1})}

\let\bar=\overbar

\def\etal{{\it et al.}}

\def\lsim{\mathrel{\raise.3ex\hbox{$<$\kern-.75em\lower1ex\hbox{$\sim$}}}}
\def\gsim{\mathrel{\raise.3ex\hbox{$>$\kern-.75em\lower1ex\hbox{$\sim$}}}}

\def\L{{\cal L}}

\def\L{{\cal L}}

\def\half{\frac{1}{2}}

\def\del{\partial}
\def\Dslash{\not{\hbox{\kern-4pt $D$}}}
\def\dslash{\not{\hbox{\kern-2pt $\del$}}}

\def\Dlr{\mathrel{\raise1.5ex\hbox{$\leftrightarrow$\kern-1em\lower1.5ex\hbox{$D$}}}}

\def\ee{e^+e^-}

\def\msb{{\bar{\scriptsize M \kern -1pt S}}}

\def\drb{{\bar{\scriptsize D \kern -1pt R}}}

\def\s#1{\widetilde{#1}}

\makeatletter
\def\section{\@startsection{section}{0}{\z@}{5.5ex plus .5ex minus
 1.5ex}{2.3ex plus .2ex}{\large\bf}}
\def\subsection{\@startsection{subsection}{1}{\z@}{3.5ex plus .5ex minus
 1.5ex}{1.3ex plus .2ex}{\normalsize\bf}}
\def\subsubsection{\@startsection{subsubsection}{2}{\z@}{-3.5ex plus
-1ex minus  -.2ex}{2.3ex plus .2ex}{\normalsize\sl}}

\renewcommand{\@makecaption}[2]{%
   \vskip 10pt
   \setbox\@tempboxa\hbox{\small #1: #2}
   \ifdim \wd\@tempboxa >\hsize     
       \small #1: #2\par          
     \else                        
       \hbox to\hsize{\hfil\box\@tempboxa\hfil}
   \fi}

\makeatother

%% file: NewHiggsv3.bbl
\begin{thebibliography}{99}
\bibitem{gapology}
In the description of Higgs physics by 
dimension-6 operators, the new physics corrections to
the $hgg$ couplings actually depend on two EFT coefficients $c_{Ht}$
and $c_{GG}$, representing the  corrections to the $hgg$ loop diagram
from corrections to the $ht\bar t$
coupling and from new heavy quarks. 
These corrections cannot be separated using data from reactions in
which the Higgs remains on-shell, so we summarize these effects here
in a single parameter called $c_{Hg}$.
\bibitem{detsix}
Determination of $c_6$ in this model-independent EFT framework is
discussed in \cite{BFJPT}. This 
requires $\ee$ collisions of at least 500~GeV.
\bibitem{CLICdisclaimer}
The actual program presented in \cite{CLIC} would take only
500~fb$^{-1}$ of data a 380~GeV and then upgrade to a center of mass
energy of 1.4~TeV.  One might question the validity of an EFT analysis
at 1.4~TeV, and we have not studied this issue  sufficienty.  We hope
that the comparison shown  in Table~\ref{tab:polfits} is sufficiently
informative.
\bibitem{Wfrompol}
The errors estimated in the Appendix for the new physics parameters in
$\ee\to W^+W^-$ make use of the polarization asymmetry, among other
observables.  For simplicity, we input the same errors in our fits for 
unpolarized colliders.
\bibitem{sigmaex}
Technically, we convert the $\chi^2$ for model A given model B to a $p$-value, then
quote this as a number of $\sigma$s for a 1-degree-of freedom
comparison.  The result is very close to $[\chi^2]^{1/2}$.
\end{thebibliography}
